\documentclass[pdftex,twocolumn,epjc3]{svjour3}          

\RequirePackage[T1]{fontenc}

\smartqed  

\RequirePackage{graphicx}
\RequirePackage{mathptmx}     
\usepackage{float}
\usepackage{tensor}
\usepackage{braket}
\usepackage{color}
\RequirePackage{flushend}
\RequirePackage[numbers,sort&compress]{natbib}
\RequirePackage[colorlinks,citecolor=blue,urlcolor=blue,linkcolor=blue]{hyperref}
\usepackage{multirow}
\usepackage{graphicx}
\journalname{Eur. Phys. J. C}
\begin{document}
\title{Warm dense matter and cooling of supernovae remnants}
\author{Ankit Kumar\thanksref{e1,addr1,addr2}
        \and
        H. C. Das\thanksref{e2,addr1,addr2}
        \and
        S. K. Biswal \thanksref{addr3}
        \and
        Bharat Kumar \thanksref{addr4,addr5}
        \and
        S. K. Patra \thanksref{e3,addr1,addr2}
}
\thankstext{e1}{e-mail: ankit.k@iopb.res.in}
\thankstext{e2}{e-mail: harish.d@iopb.res.in}
\thankstext{e3}{e-mail: patra@iopb.res.in}
\institute{Institute of Physics, Sachivalaya Marg, Bhubaneswar-751005, India \label{addr1}
\and
Homi Bhabha National Institute, Training School Complex, Anushakti Nagar, Mumbai 400085, India \label{addr2}
\and 
Department of Astronomy, Xiamen University, Xiamen 361005, P. R. China \label{addr3}
\and
Department of Physics $\&$ Astronomy, National Institute of Technology, Rourkela, India \label{addr4}
\and
Center for Computational Sciences, University of Tsukuba, Tsukuba 305-8577, Japan\label{addr5}
}
\date{Received: xxxx / Accepted: xxxx}
\maketitle
\begin{abstract}
We study the thermal effects on the nuclear matter (NM) properties such as binding energy, incompressibility, free symmetry energy and its coefficients using NL3, G3 and IU-FSU parameter sets of relativistic mean-field models. These models being consistent with the properties of cold NM, have also been used to study the effect of temperature by incorporating the Fermi function. The critical temperature for the liquid-gas phase transition in the symmetric NM is found to be 14.60, 15.37 and 14.50 MeV for NL3, G3 and IU-FSU parameter sets respectively, which is in excellent agreement with previous theoretical and experimental studies. We inspect that the properties related to second differential coefficient of the binding energy and free symmetry energy at saturation density ( i.e. $K_{0}(n,T)$ and $Q_{sym,0}$ ) exhibit the contrary effects for NL3 and G3 parameters as the temperature increases. We find that the prediction of saturated curvature parameter ( $K_{sym,0}$ ) for G3 equation of state at finite temperature favour the combined analysis of $K_{sym,0}$ for the existence of massive pulsars, gravitational waves from GW170817 and NICER observations of PSR J0030+0451. Further, we investigate the cooling mechanism of newly born stars through neutrino emissivity controlled by direct Urca process and instate some interesting remarks about neutrino emissivity. We also deliberate the effect of temperature on the M-R profile of Proto-Neutron star.
\end{abstract}
\section{Introduction}
\label{intro}
One of the most prominent energetic events of the universe is manifested by the core-collapse supernovae (CCSN) explosion of the giant stars having mass in the range of 8 - 40 times that of the mass of sun \cite{Gilmore1915}. The concept that the gravitational instability during the evolution of a massive star results in a rapid compression and then thermonuclear explosion, was first suggested by Burbidge, Burbidge, Fowler, and Hoyle (designated as "$B^2FH$") \cite{Colgate}. The energy emerge during this explosion is carried off by the photons and the neutrinos, which are billion trillion trillion in numbers and hauled the most of the energy released. The observation of Einstein Observatory (HEAO-2) first supported the fact that high neutrino emissivity is mainly responsible for the rapid cooling of newly born dense star \cite{Boguta}. Studies suggest that the neutrinos also play a significant role in the dynamics of supernovae explosion and control many important aspects of the collapsed core, generally known as newly born neutron star \cite{Burrows, Barbara}. In this exploring era of every tiny and massive object through modern science, neutron stars still hold the mystery in itself and are poorly known objects. These are the ideal play-field in the observable universe to inspect the theories of dense matter physics and unfold new opportunities. A large part of our perception about the dynamics of supernovae explosion and the composition of neutron stars is based on the equation of state (EoS), which makes the EoS a key ingredient of our study. Since it is believed that the matter evolved during the core-collapse phenomenon is unable to attain $\beta-$equilibrium condition for few seconds \cite{Stone}, due to the promptness of the CCSN explosion, so, usually the EoS for a dense matter with either same amount of protons and neutrons or a proton fraction of $\sim 0.2$ is taken into account to cipher the dynamics of collapse event \cite{Mezzacappa}. Many other important aspects of core-collapse events like how many protons are converted into neutrons, quenching rate, the mass-radius profile of newly born neutron star and its composition are determined by the EoS \cite{Schneider}.

Another important dimension which makes the EoS of nuclear matter (NM) at finite temperature more interesting is the study of the dynamics of heavy ion collision reaction and structure of exotic nuclei. Under extreme conditions of density, isospin and temperature the EoS can be explored through many experimental facilities such as GANIL
-SPIRAL2 \cite{France} facility in France, CSR in China \cite{ZHIYU} and the FRIB technique in the United States \cite{US}. Nuclear collisions can produce hot and dense hadronic NM in terrestrial laboratories momentarily, which can procure a lot of information about the thermodynamical properties of NM like incompressibility, symmetry energy and its derivatives \cite{Danielewicz}. In heavy-ion collision experiments, the symmetry energy plays an important role in understanding the isoscaling behaviour of multi-fragmentation phenomena. The observations of the experiments at Texas A$\&$M University and National Superconducting Cyclotron Laboratory  suggest a connection between the nuclear symmetry energy and primary fragment yield distribution of statistical multi-fragmentation model \cite{AkiraOno,Souliotis,Iglio}. On the other hand, symmetry energy at finite temperature also plays an important role in the cooling mechanism of newly born hot astrophysical objects \cite{Prakash,Takatsuka,Bao-AnLi} and has effective impact on the cooling rate through direct Urca and modified Urca processes \cite{717Y,refId0}.

In the recent years, the density dependence of the symmetry energy and its derivatives ($L_{sym}$, $K_{sym}$, $Q_{sym}$) have been used to constrain the EoS near saturation density, which make them more vital to decipher properly. It is well known that the slope parameter ($L_{sym}$) and the size of the neutron skin in super-heavy nuclei are connected by a strong linear correlation. The slope ($L_{sym}$) and curvature parameter ($K_{sym}$) also control the location of the neutron drip line, core-crust transition density and gravitational binding energy of neutron star \cite{Tsang2}. The isovector skewness parameter ($Q_{sym}$) is the most ambiguous quantity, due to the large fluctuations in its value obtained from different models. Recent studies show that $Q_{sym}$ is related to the incompressibility \cite{Typel} of the system and also suggest an important role of $Q_{sym}$ in the cooling of newly born proto-neutron star \cite{Schneider}. We study in detail the effects of temperature on all these nuclear parameters and the correlations among them using the most familiar NL3, IU-FSU and recently developed G3 parameter sets. With the thermal effects being the main focus of our study, we also explore the dependence of thermal index ($\Gamma_{th}$) on the density for all the parameter sets.
In the extant work, we also investigate the dependence of cooling mechanism of a hot dense matter on the EoS and the variation in the mass-radius profile of a proto-neutron star with temperature. The temperature in the interior of a newly born dense star just after the supernova explosion can vary from 10 to 100 MeV \cite{Oechslin}. The analyzation of the x-rays emitted during the stellar evolution of the young star ensure the fact that it losses most of its energy through extremely rapid neutrino emission enhanced by the direct Urca process (named after a casino in Rio) \cite{Brown,Kaminker,Yakovlev}. However, the direct Urca process dominates only during the initial stage of the cooling and can no longer operate once the proton fraction inside the core reaches a threshold value. The proton fraction in a newly born proto-neutron star depends on how the nuclear symmetry energy scales with density -- which is an important aspect of this work \cite{Brown}. In the mean-field approximation, the symmetry energy and the magnitude of the neutrino emissivity $(Q)$ seem to be controlled by the EoS of different parameter sets.

The  paper  is  organized  as: In section \ref{QHD}, we explain the formalism for temperature dependent quantum hadron-
dynamics model. Section \ref{SNM} is divided into two parts, where \ref{SNM frame} is devoted to the theoretical structure of the relativistic mean-field formalism at finite temperature and acquaintance with the NM parameters. In sub-section \ref{SNM results} we presented the results for various NM parameters of symmetric NM using NL3, G3 and IU-FSU parameter sets. The detailed framework and the results of neutrino emissivity through direct Urca process are extended in section \ref{cooling}. The mass-radius profile of the proto-neutron star is discussed in section \ref{star}. Finally, the discussion and the concluding remarks are outlined in the section \ref{discussion}.   
\section{Temperature Dependent QHD Model}\label{QHD}
Temperature dependent QHD model is based on relativistic covariant field theory, proposed to obtain the expediential in-medium nuclear properties around the saturation density. The nucleons (neutrons and protons) are represented by the Dirac spinor $\psi$ in the Lagrangian density. We take in account the $\sigma$, $\omega$, $\rho$ and $\delta$ mesons in our Lagrangian, represented by the $\sigma$, $\omega^\mu$, $\vec\rho^{\,\mu}$ and $\vec\delta$ fields respectively. The mesons act as the mediators and represent the effective nuclear interaction of the nucleons through the meson fields $\sigma$, $\omega^\mu$, $\vec\rho^{\,\mu}$ and $\vec\delta$. The basic Lagrangian density used in the present work, which includes the interaction of nucleon fields $\psi$ with a scalar field ($\sigma$), a vector field ($\omega$) and isovector fields ($\vec\rho^{\,\mu}$ and $\vec\delta$) and the cross-coupled interactions of these meson fields up to fourth order \cite{S.K.Singh}, is given by
%
\begin{eqnarray}\label{lag}
{\cal L} & = &  \sum_{\alpha=p,n} \bar\psi_{\alpha}
\Bigg\{\gamma_{\mu}(i\partial^{\mu}-g_{\omega}\omega^{\mu}-\frac{1}{2}g_{\rho}\vec{\tau}_{\alpha}\!\cdot\!\vec{\rho}^{\,\mu})
-(M-g_{\sigma}\sigma\nonumber\\
&&
-g_{\delta}\vec{\tau}_{\alpha}\!\cdot\!\vec{\delta})\Bigg\} \psi_{\alpha}
  +\frac{1}{2}
   \partial^{\mu}\sigma\,\partial_{\mu}\sigma
  -\frac{1}{2}m_{\sigma}^{2}\sigma^2+\frac{\zeta_0}{4!}g_\omega^2
   (\omega^{\mu}\omega_{\mu})^2
   \nonumber \\
& & \null 
-g_{\sigma}\frac{m_{\sigma}^2}{M}
\Bigg(\frac{\kappa_3}{3!}
  + \frac{\kappa_4}{4!}\frac{g_{\sigma}}{M}\sigma\Bigg)
   \sigma^3
+\frac{1}{2}m_{\omega}^{2}\omega^{\mu}\omega_{\mu}
   -\frac{1}{4}F^{\mu\nu}F_{\mu\nu}\nonumber\\
 &&  \null
   +\frac{1}{2}\frac{g_{\sigma}\sigma}{M}\Bigg(\eta_1+
 \frac{\eta_2}{2} \frac{g_{\sigma}\sigma}{M}\Bigg)m_\omega^2\omega^{\mu}\omega_{\mu}+\frac{1}{2}\eta_{\rho}\frac{m_{\rho}^2}{M}g_{\sigma}\sigma(\vec\rho^{\,\mu}\!\cdot\!\vec\rho_{\mu}) \nonumber\\
 &&
 +\frac{1}{2}m_{\rho}^{2}\rho^{\mu}\!\cdot\!\rho_{\mu} -\frac{1}{4}\vec R^{\mu\nu}\!\cdot\!\vec R_{\mu\nu}
  -\Lambda_{\omega}g_{\omega}^2g_{\rho}^2(\omega^{\mu}\omega_{\mu})(\vec\rho^{\,\mu}\!\cdot\!\vec\rho_{\mu})
 \nonumber\\
 &&
 +\frac{1}{2}\partial^{\mu}\vec\delta\,\partial_{\mu}\vec\delta-\frac{1}{2}m_{\delta}^{2}\vec\delta^{\,2},
\label{eq1}
\end{eqnarray}
%
where M is the nucleon mass; $m_\sigma$, $m_\omega$, $m_\rho$ and $m_\delta$ are the masses of mesons and $g_\sigma$, $g_\omega$, $g_\rho$ and $g_\delta$ are the coupling constants for the $\sigma$, $\omega$, $\rho$ and $\delta$ mesons respectively; $\kappa_3$ (or $\kappa_4$) and $\zeta_0$ are the self-interacting coupling constants of the $\sigma$ and $\omega$ mesons respectively; $\eta_1$, $\eta_2$, $\eta_\rho$ and $\Lambda_\omega$ are the coupling constants of non-linear cross-coupled terms. The quantities $F^{\mu\nu}$ and $\vec R^{\mu\nu}$ being the field strength tensors for the $\omega$ and $\rho$ mesons respectively, defined as $F^{\mu\nu}$ = $\partial^\mu\omega^\nu-\partial^\nu\omega^\mu$ and $\vec R^{\mu\nu}$ = $\partial^\mu\vec\rho^{\,\nu}-\partial^\nu\vec\rho^{\,\mu}$. The $\vec\tau$ are the Pauli matrices and behave as the isospin operator, which carry the isospin component of the nucleons.
In the further calculations of the present work, we will consider relativistic mean-field approximation and isotropy in isospin space. Within the assumed  approximation, the meson fields are replaced by their expectation values and for rotationally invariant systems only the time component $(\mu=0)$ of the isovector field survives \cite{Aguirre, Liu_Greco}. The third component of the isospin operator ($\tau_3$) when operates on neutron and proton gives, $\tau_3\ket{p}=(+1)\ket{p}$ and $\tau_3\ket{n}=(-1)\ket{n}$. Applying the relativistic mean-field approximation, the Lagrangian density yields the following Dirac equation for the nucleon field
%
\begin{eqnarray}
\Bigg\{i\gamma_{\mu}\partial^{\mu}-g_{\omega}\gamma_{0}\omega-\frac{1}{2}g_{\rho}\gamma_{0}\tau_{3\alpha}\rho-M^\ast_\alpha\Bigg\} \psi_\alpha=0,
\end{eqnarray}
%
where $M^\ast_\alpha$ is the effective mass of the nucleon. Taking into account the third component of the isospin for neutron and proton as defined above, the effective masses of the neutron and proton are given by
%
\begin{eqnarray}
M_{p}^{\ast}&=&M-g_{s}\sigma-g_{\delta}\delta,  \\
M_{n}^{\ast}&=&M-g_{s}\sigma+g_{\delta}\delta,   
\end{eqnarray}
%
The Lagrangian density can be used to derive  Euler-Lagrange equations of motion for the meson fields with the help of relativistic mean-field approximation as \cite{Bharat}
%
\begin{eqnarray}
m_{\sigma}^2\sigma & = & g_{\sigma} \sum_{\alpha=p,n}\langle\bar\psi_\alpha\gamma_0\psi_\alpha\rangle - \frac{m_\sigma^2g_\sigma}{M}\sigma^2\Bigg(\frac{\kappa_3}{2} + \frac{\kappa_4}{3!}\frac{g_\sigma\sigma}{M}\Bigg)\nonumber\\
&&
+\frac{g_\sigma}{2M}\Bigg(\eta_1 + \eta_2\frac{g_\sigma\sigma}{M}\Bigg){ m_\omega^2\omega^2} + \frac{\eta_{\rho}}{2M}{g_\sigma}{m_\rho^2 }{\rho^2}, \\
%
m_\omega^2\omega & = & g_\omega \sum_{\alpha=p,n} \langle\bar\psi_\alpha\psi_\alpha\rangle - \Bigg(\eta_1 + \frac{\eta_2}{2}\frac{g_\sigma\sigma}{M}\Bigg)\frac{g_\sigma\sigma}{M}m_\omega^2\omega\nonumber\\
&& - \frac{1}{3!}\zeta_0g_{\omega}^2\omega^3 - 2\;\Lambda_{\omega} g_{\omega}^2 g_{\rho}^2 \rho^2 \omega, \\
%
m_{\rho}^2 \rho & = &\frac{1}{2}g_{\rho} \sum_{\alpha=p,n} \langle\bar\psi_\alpha\tau_3\psi_\alpha\rangle - \eta_\rho \frac{g_\sigma\sigma}{M}m_{\rho}^2\rho \nonumber\\
&&
- 2\;\Lambda_{\omega} g_\omega^2 g_\rho^2 \omega^2 \rho,\\
%
m_\delta^2 \delta & = & g_{\delta} \sum_{\alpha=p,n} \langle\bar\psi_\alpha\tau_3\gamma_0\psi_\alpha\rangle. 
\end{eqnarray}
Assuming NM as a uniform system, where the fields of $\sigma$, $\omega$ and $\rho$ mesons are independent of position and taking the thermodynamic argument of free gas into consideration, the baryon, scalar and isovector densities for the appropriate Fermi momentum at finite temperature can be evaluated as \cite{Wang,Yang,Bharat}
\begin{eqnarray}
n &=& \sum_{\alpha=p,n} \langle\bar\psi_\alpha\psi_\alpha\rangle  = n_{p}+n_{n} \nonumber\\
&&
= \sum_{\alpha=p,n}\frac{2}{(2\pi)^{3}}\int_{0}^{k_{\alpha}}d^{3}k\,[f_{\alpha}(\mu_{\alpha}^{\ast},T)-\bar f_{\alpha}(\mu_{\alpha}^{\ast},T)], \\ 
n_s &=& \sum_{\alpha=p,n}\langle\bar\psi_\alpha\gamma_0\psi_\alpha\rangle = n_{sp}+n_{sn} \nonumber\\
&&
= \sum_{\alpha=p,n} \frac{2}{(2\pi)^3}\int_{0}^{k_{\alpha}} d^{3}k\, 
\frac{M_{\alpha}^{\ast}} {(k^{2}_{\alpha}+M_{\alpha}^{\ast 2})^{\frac{1}{2}}}
\Big[f_{\alpha}(\mu_{\alpha}^{\ast},T)
\nonumber\\
 &&
+\bar f_{\alpha}(\mu_{\alpha}^{\ast},T)\Big], \\ 
n_3 &=& \sum_{\alpha=p,n} \langle\bar\psi_\alpha\tau_3\psi_\alpha\rangle = n_p-n_n, \\ 
n_{s3} &=& \sum_{\alpha=p,n} \langle\bar\psi_\alpha\tau_3\gamma_0\psi_\alpha\rangle = n_{sp}-n_{sn},
\end{eqnarray}
where $n_p$ and $n_n$ are the proton and neutron densities respectively, $k_\alpha$ is the nucleon Fermi momentum, T is the temperature, $f_{\alpha}(\mu_{\alpha}^{\ast},T)$ and $\bar f_{\alpha}(\mu_{\alpha}^{\ast},T)$ are thermal Fermi distribution function of the nucleon and the anti-nucleon, and $\mu_\alpha^{\ast}$ is the effective chemical potential of the nucleon. The familiar Fermi distribution function for the particle and the anti-particle can be written as \cite{Wang} 
%
\begin{eqnarray}
f_{\alpha}(\mu_{\alpha}^{\ast},T)&=&\frac{1}{e^{({\cal E}_{\alpha}^{\ast}-\mu_{\alpha}^{\ast})/k_B T}+1}, \\ \nonumber \\ 
\bar{f_{\alpha}}(\mu_{\alpha}^{\ast},T)&=&\frac{1}{e^{({\cal E}_{\alpha}^{\ast}+\mu_{\alpha}^{\ast})/k_B T}+1},
\end{eqnarray}
%
where $k_B$ is the Boltzmann constant and ${\cal E_\alpha^\ast}$ is the effective energy of the nucleon which can be written as
%
\begin{eqnarray}
{\cal E_\alpha^\ast} = \sqrt{k_\alpha^2+M_\alpha^{\ast2}}.
\end{eqnarray}
%
The effective chemical potential of the proton and the neutron can be derived as \cite{Wang}
%
\begin{eqnarray}
\mu_{p}^{\ast}&=&\mu_{p}-g_{\omega}\omega-\frac{1}{2}g_{\rho}\rho,   \\ \nonumber \\
\mu_{n}^{\ast}&=&\mu_{n}-g_{\omega}\omega+\frac{1}{2}g_{\rho}\rho,
\end{eqnarray}
where $\mu_p$ and $\mu_n$ are the usual the chemical potential of proton and neutron relative to free nucleon mass. Another important thing we would like to mention for the formalism is the contribution of loop correction for the renormalized theory. Recently, M. Prakash et. al. \cite{Xilin} reported that the two-loop approximation improves the energy density functional by adding density-dependent contributions to the Hartree terms of mean-field theory from the exchange of isoscalar, isovector and pseudoscalar mesons. It is found that the thermal properties of proto-neutron stars differ significantly if we include the two-loop correction in the mean-field theory. However, it has also been reported in the references ( \cite{Frun1989}, \cite{Serot1995} and \cite{Hu_2007}) that the loop contribution in the mean-field theory can be mocked up by refitting the parameters using experimentally available data. We calculate the thermal properties with the mean-field approach without loop correction because in our work the coupling constants of the Lagrangian are obtained by fitting the experimental and observational data.
%
%
\section{Symmetric Nuclear Matter}\label{SNM}
\subsection{Theoretical Formalism}\label{SNM frame}
%
In the following section, the formalism used for the calculations of symmetric and asymmetric NM properties has been discussed. To describe the asymmetric NM, we introduce the asymmetric parameter $t$, which is defined as,
\begin{eqnarray}
t = \frac{n_n-n_p}{n_n+n_p}.
\end{eqnarray}
%
We can assign the desired asymmetry in the NM by varying the value of $t$. For symmetric NM (SNM), $t=0$ and $t=1$ for pure neutron matter. In the present work, we extend our calculations for SNM ($t=0$). The energy density and the pressure for the NM can be calculated from the Lagrangian using the expression for energy-momentum tensor \cite{Walecka}, which is
%
\begin{eqnarray}
T^{\mu\nu} =  \sum_{j} \frac{\partial {\cal L}}{\partial(\partial_\mu\phi_j)}\partial^{\nu}\phi_j - \eta^{\mu\nu}{\cal L}, 
\end{eqnarray}
%
where $\phi_j$ includes all the fields present in the Lagrangian. Using this expression of the energy density and pressure for a warm nuclear system can be naively derived as \cite{S.K.Singh}
%
\begin{eqnarray}\label{ed}
E & = & \sum_{\alpha=p,n} \frac{2}{(2\pi)^{3}}\int_{0}^{k_{\alpha}} d^{3}k\, {\cal E}_{\alpha}^\ast (k) \Big[f_{\alpha}(\mu_{\alpha}^{\ast},T)+\bar f_{\alpha}(\mu_{\alpha}^{\ast},T)\Big] 
\nonumber\\
&&
+n\,g_\omega\,\omega+m_{\sigma}^2{\sigma}^2\Bigg(\frac{1}{2}+\frac{\kappa_{3}}{3!}\frac{g_\sigma\sigma}{M}+\frac{\kappa_4}{4!}\frac{g_\sigma^2\sigma^2}{M^2}\Bigg)-\frac{1}{4!}\zeta_{0}\,{g_{\omega}^2}\,\omega^4
\nonumber\\
&&
 -\frac{1}{2}m_{\omega}^2\,\omega^2\Bigg(1+\eta_{1}\frac{g_\sigma\sigma}{M}+\frac{\eta_{2}}{2}\frac{g_\sigma^2\sigma^2}{M^2}\Bigg)+ \frac{1}{2}n_{3}\,g_\rho\,\rho \nonumber\\
 && 
  - \frac{1}{2}\Bigg(1+\frac{\eta_{\rho}g_\sigma\sigma}{M}\Bigg)m_{\rho}^2\,\rho^{2}-\Lambda_{\omega}\, g_\rho^2\, g_\omega^2\, \rho^2\, \omega^2
+\frac{1}{2}m_{\delta}^2\, \delta^{2},
\\
%
and \nonumber \\ \nonumber
P & = & \sum_{\alpha=p,n} \frac{2}{3 (2\pi)^{3}}\int_{0}^{k_{\alpha}} d^{3}k\, \frac{k^2}{{\cal E}_{\alpha}^\ast(k)} \Big[f_{\alpha}(\mu_{\alpha}^{\ast},T)+\bar f_{\alpha}(\mu_{\alpha}^{\ast},T)\Big]\nonumber\\
&& - m_{\sigma}^2{\sigma}^2\Bigg(\frac{1}{2} + \frac{\kappa_{3}}{3!}\frac{g_\sigma\sigma}{M} + \frac{\kappa_4}{4!}\frac{g_\sigma^2\sigma^2}{M^2}\Bigg)+ \frac{1}{4!}\zeta_{0}\,{g_{\omega}^2}\,\omega^4 
\nonumber\\
& &
 +\frac{1}{2}m_{\omega}^2\,\omega^2\Bigg(1+\eta_{1}\frac{g_\sigma\sigma}{M}+\frac{\eta_{2}}{2}\frac{g_\sigma^2\sigma^2}{M^2}\Bigg)+\Lambda_{\omega}\, g_\rho^2\, g_\omega^2\, \rho^2\, \omega^2
  \nonumber\\
& &
+ \frac{1}{2}\Bigg(1+\frac{\eta_{\rho}g_\sigma\sigma}{M}\Bigg)m_{\rho}^2\,\rho^{2}-\frac{1}{2}m_{\delta}^2\, \delta^{2}.
\end{eqnarray}
%
Since we deal with the temperature dependent NM, we also did some study regarding phase coexistence, which unfolds the liquid-gas phase transition in thermal NM. The NM system can remain only in one phase (gaseous) once it reaches the critical point \cite{Wang}. For symmetric NM, the inflection point of the pressure curve with respect to the total nucleon density determines the critical point \cite{Yang}, that is
\begin{eqnarray}\label{critical}
\frac{\partial P}{\partial n}\Bigg |_{T=T_C} = \frac{\partial^2P}{\partial n^2}\Bigg |_{T=T_C} = 0,
\end{eqnarray}
$T_{C}$ being the critical temperature. One of the basic and fundamental quantity of the NM is incompressibility, also known as the isoscalar incompressibility or compression modulus $(K)$ \cite{Schneider}. $K$ directly influences the curvature of the equation of state and gives adequate information about the nature of the equation of state. Higher the value of K, more stiff the equation of state will be \cite{NgoHaiTan}. The incompressibility is related to the equation of state through \cite{Walecka}
\begin{eqnarray}
K (n,T) &=& 9\,n^2 \, \frac{\partial ^2 (E/n)}{\partial n^2}.
\end{eqnarray}
Another important quantity which controls the equation of state of NM is symmetry energy. Symmetry energy has a significant contribution to the pressure of the astrophysical objects which is responsible for gravitational attraction \cite{Goudarzi}. In case of the hot NM, we can define two forms of symmetry energy, one we call as the nuclear symmetry energy $(E_{sym})$ and the other as free symmetry energy $(F_{sym})$. To calculate the free symmetry energy of the hot NM, we need entropy density $S$ and free energy density $F$ of that system. So the free energy density is given by
\begin{eqnarray}
F &=& E-TS, 
\end{eqnarray}
and the entropy density for a hot NM is calculated by \cite{Walecka}
\begin{eqnarray}
S &=& - \sum_{\alpha=p,n} \frac{2}{(2\pi)^{3}}\int_{0}^{k_{\alpha}} d^{3}k\, \Bigg[f_{\alpha}(\mu_{\alpha}^{\ast},T) \ln{f_{\alpha}(\mu_{\alpha}^{\ast},T)} \nonumber\\
&& 
+ (1-f_{\alpha}(\mu_{\alpha}^{\ast},T)) \ln{(1-f_{\alpha}(\mu_{\alpha}^{\ast},T))}
+ \bar f_{\alpha}(\mu_{\alpha}^{\ast},T)\nonumber\\
&& 
\ln{\bar f_{\alpha}(\mu_{\alpha}^{\ast},T)}
+ (1-\bar f_{\alpha}(\mu_{\alpha}^{\ast},T)) \ln{(1-\bar f_{\alpha}(\mu_{\alpha}^{\ast},T))} \Bigg].
\end{eqnarray}
Various theoretical studies have shown that we can calculate the temperature and density dependence of nuclear symmetry energy in hot symmetric NM using empirical parabolic approximation, which can be estimated as the difference of the energy per nucleon of pure neutron matter and symmetric NM \cite{NgoHaiTan}.
In a similar way, we can find the free symmetric energy for a hot NM, which is of utmost importance for astrophysical phenomena. The free symmetric energy of a symmetric NM can be defined as \cite{NgoHaiTan,Bao-AnLi}
\begin{eqnarray}
F_{sym}(n,T) = \frac{F(n,T,t=1)}{n} - \frac{F(n,T,t=0)}{n},
\end{eqnarray}
which can be interpreted as the remnant of free energy per nucleon for pure neutron matter and symmetric NM.
The free symmetry energy can be expanded in a Taylor series expansion around $\chi$. It is a dimensionless variable which describes the density deviation from saturation density and for mathematical convention and simplification of the expressions of higher order derivatives defined as, $\chi = ({n-n_0})/{3n_0}$, $n_0$ being the saturation density. So the expansion can be written as, 
\begin{eqnarray}
F_{sym}(n,T) &=& F_{sym}(n_0) + L_{sym}\,\chi + \frac{K_{sym}}{2!}\,\chi^2  \nonumber \\  &&
 + \frac{Q_{sym}}{3!}\,\chi^3 + O(\chi^4),
\end{eqnarray}
where $L_{sym}$, $K_{sym}$ and $Q_{sym}$ are the slope parameter, curvature parameter and skewness parameter respectively. Recently it is convinced that these parameters are of utmost importance in nuclear and astrophysics by showing the correlation of these parameters with different nuclear and astrophysical properties \cite{Schneider}. The expressions for these parameters can be extracted as \cite{Chen}
\begin{eqnarray}
L_{sym}(n,T) &=& 3n\,\frac{\partial F_{sym}(n,T)}{\partial n}, \\ 
K_{sym} (n,T) &=& 9n^2\,\frac{\partial^2 F_{sym}(n,T)}{\partial n^2}, \\ 
Q_{sym} (n,T) &=& 27n^3\,\frac{\partial^3 F_{sym}(n,T)}{\partial n^3}. 
\end{eqnarray}
Although we extract the results from this expansion at high densities, we should not exclude the fact that at very high nuclear densities the error bar in higher-order coefficients extracted through series expansion is quite high \cite{Schneider}. Thermal index commits an important aspect on core-collapse supernovae simulations \cite{Yasin}. The thermal index can be calculated with the help of simple formula discussed in \cite{Carbone}  
\begin{eqnarray}
\Gamma_{th} &=& 1 + \frac{E_{th}}{P_{th}}
\end{eqnarray}
where $E_{th}$ and $P_{th}$ are the thermal energy density and pressure respectively. $E_{th}$ and $P_{th}$ are given by $E_{th} = E(T) - E(0)$ and $P_{th} = P(T) - P(0)$, where $E(T)$ and $P(T)$ are the energy density and pressure at temperature T.
\subsection{Results}\label{SNM results}
\begin{figure}[H]
    \centering
    \includegraphics[width=0.8\columnwidth]{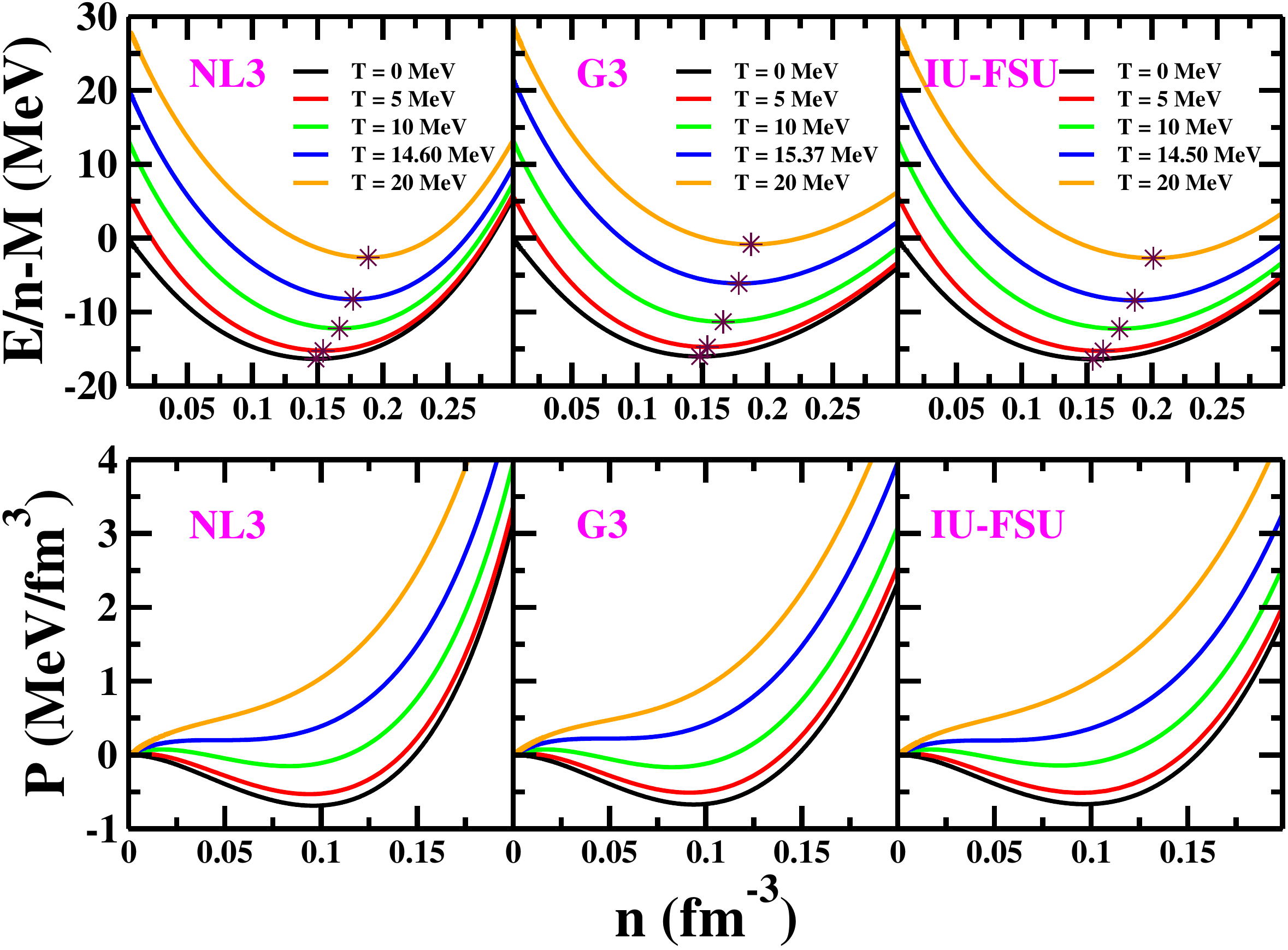}
    \caption{(colour online) Binding energy and pressure as a function of nucleon density for symmetric NM at different temperatures. The left, middle and right panels show the results for NL3, G3 and IU-FSU parameter sets respectively.}
    \label{EOS}
\end{figure}

\begin{figure}[H]
    \centering
    \includegraphics[width=0.8\columnwidth]{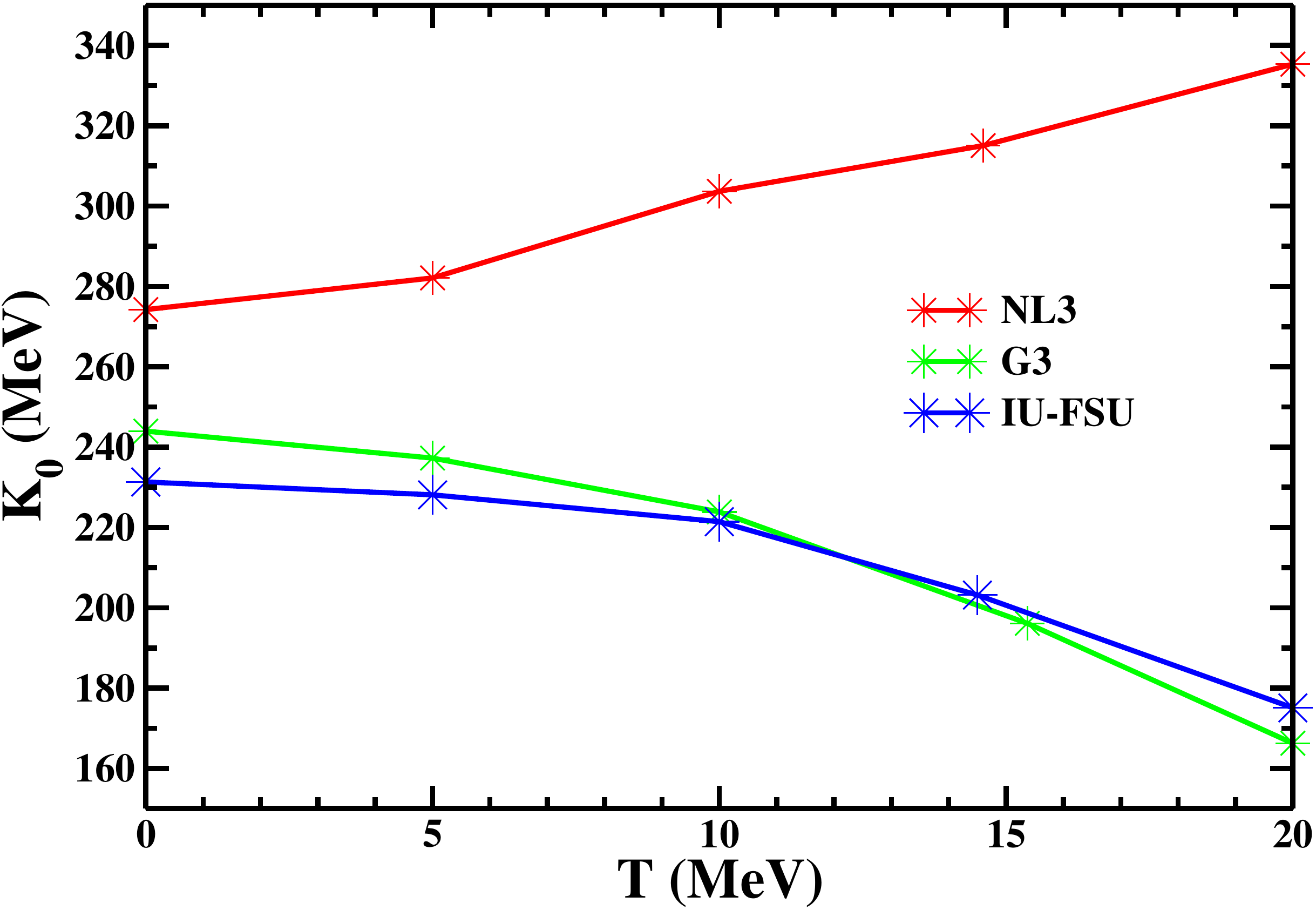}
    \caption{(colour online) Variation of incompressibility at saturation density with temperature for SNM ($t=0$).}
    \label{incompvari}
\end{figure}

\begin{figure}[H]
    \centering
    \includegraphics[width=0.8\columnwidth]{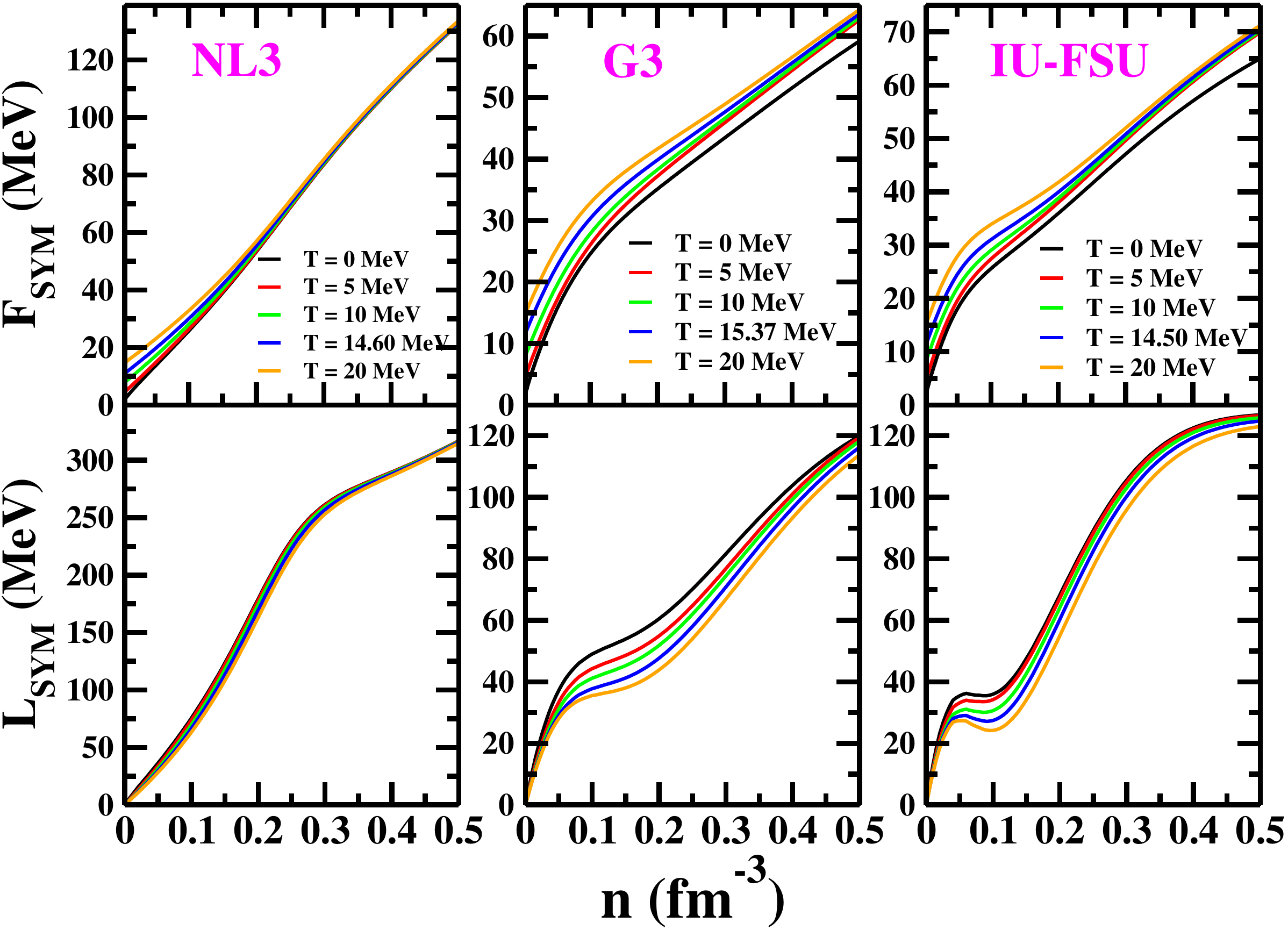}
    \caption{(colour online) Symmetry energy ($F_{sym}$) and $L_{sym}$ as a function of nucleon density for SNM ($t=0$). The left panel shows the results for NL3, middle panel for G3 and right panel for IU-FSU parameter set.}
    \label{symm}
\end{figure}

\begin{figure}[H]
    \centering
    \includegraphics[width=0.8\columnwidth]{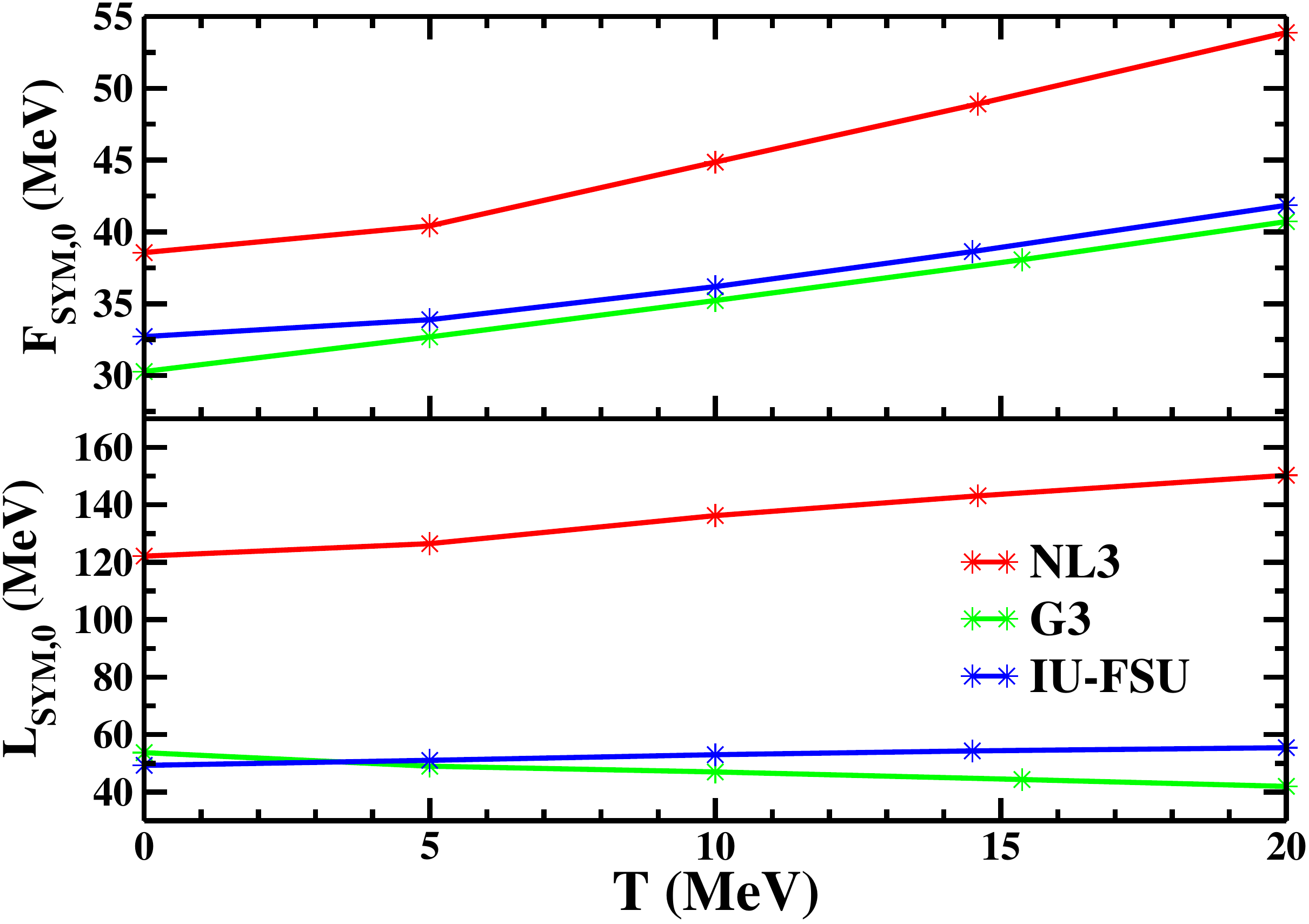}
    \caption{(colour online) Variation of symmetry energy and slope parameter at saturation density ($F_{sym,0}$ and $L_{sym,0}$) with temperature for SNM ($t=0$).}
    \label{symmvari}
\end{figure}

\begin{figure}[H]
    \centering
    \includegraphics[width=0.8\columnwidth]{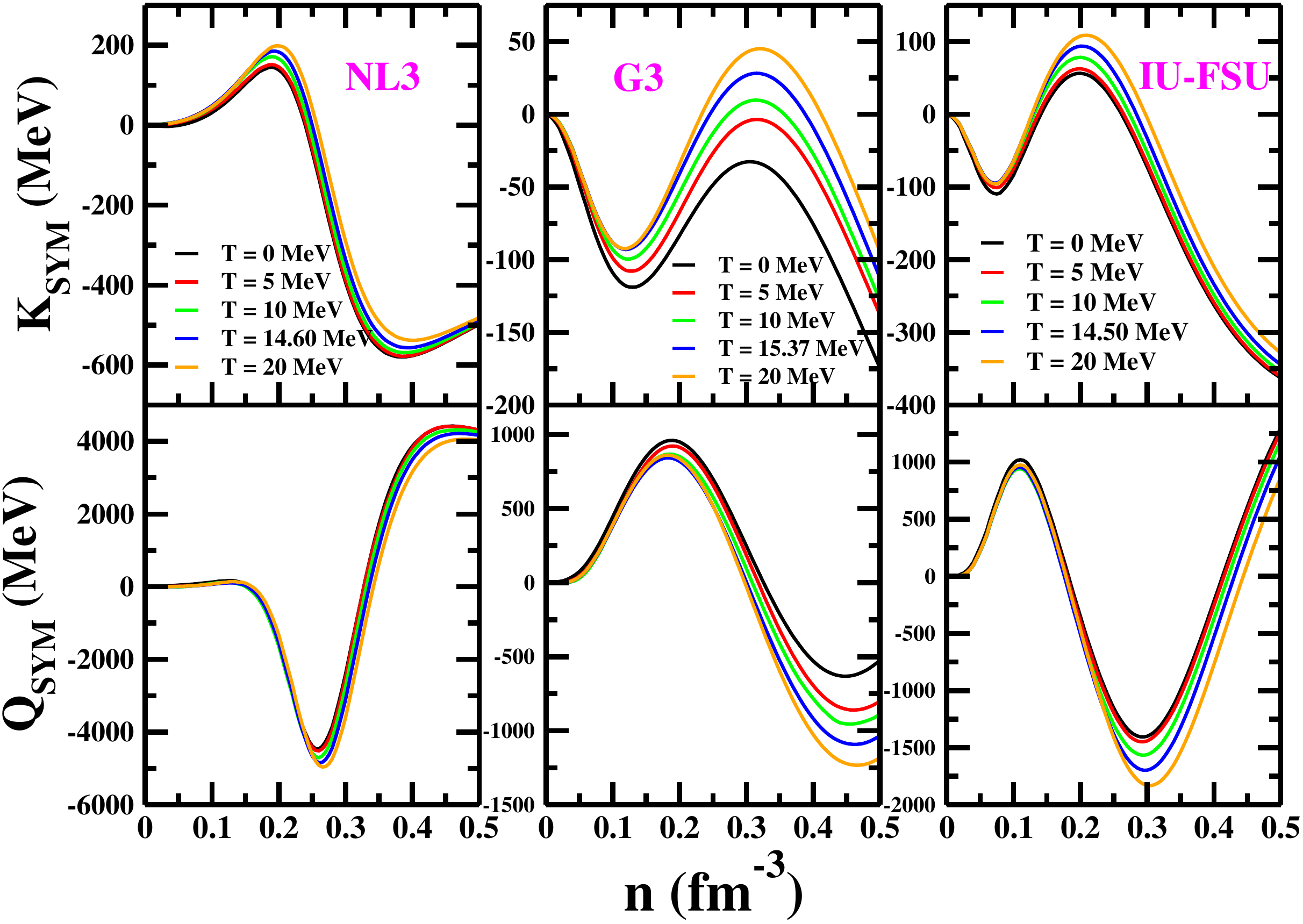}
    \caption{(colour online) $K_{sym}$ and $Q_{sym}$ as a function of nucleon density for SNM ($t=0$). The left panel shows the results for NL3 and right panel for G3 parameter set.}
    \label{Q_0}
\end{figure}

\begin{figure}[H]
    \centering
    \includegraphics[width=0.8\columnwidth]{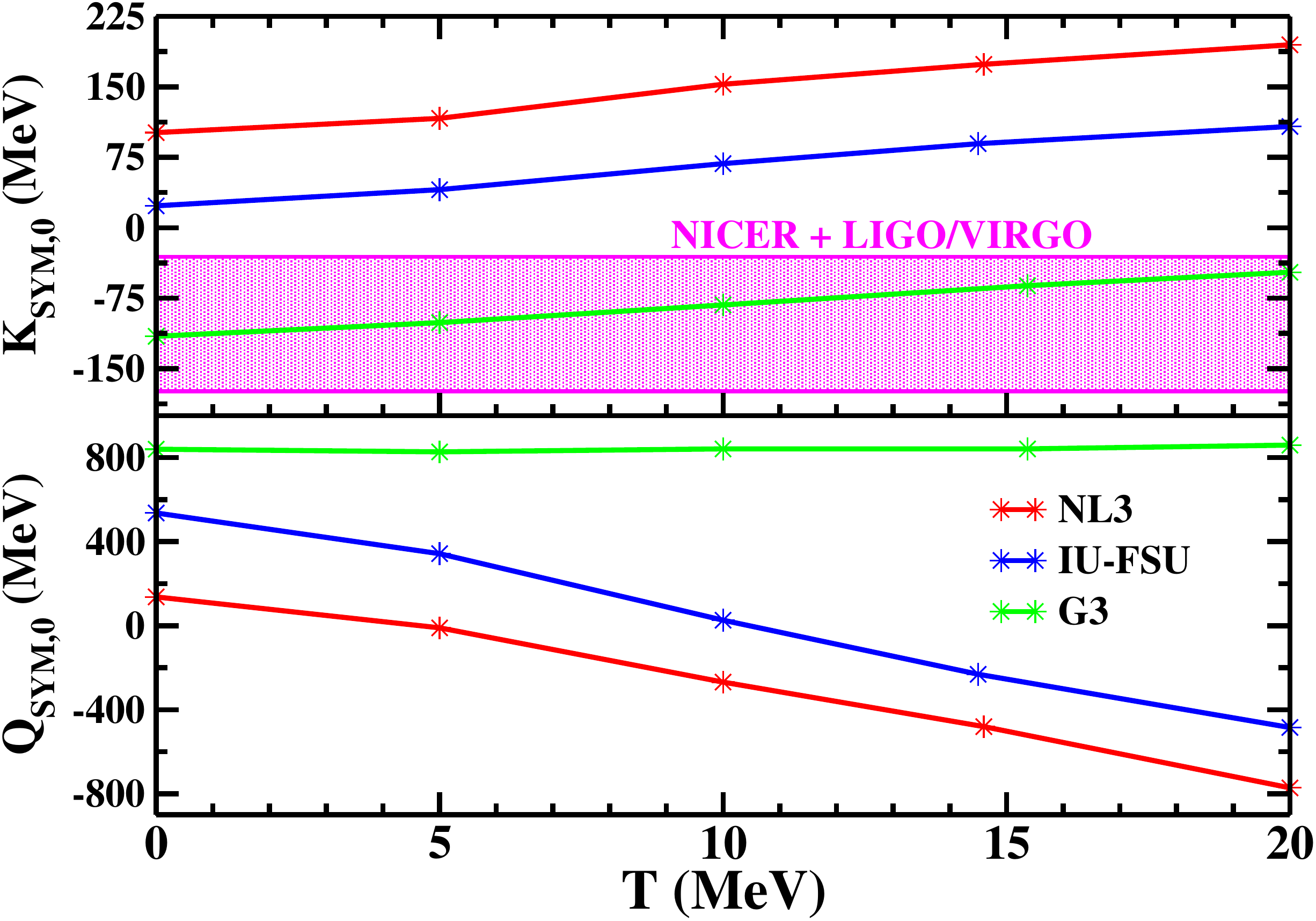}
    \caption{(colour online) Variation of Curvature parameter and Skewness parameter at saturation density ($K_{sym,0}$  and $Q_{sym,0}$ respectively) with temperature for SNM. }
    \label{Qvari}
\end{figure}

Throughout our calculations in this paper, we use NL3, G3 and IU-FSU parameter sets, where NL3 endue us the stiff EoS and the other two (G3 and IU-FSU) facilitate us to examine the softer region. The obtained results for the variation in the properties of the NM with temperature and density are discussed in the present section. The coupling constants and the empirical values of nuclear properties at saturation of cold EoS for the assumed parameter sets (NL3, G3 and IU-FSU) are given in Table \ref{table1} \cite{Lalazissis,KUMAR2017197}.
\begin{table}
\caption{The coupling constants and the NM properties at saturation for cold EoS of NL3 \cite{Lalazissis}, G3 \cite{KUMAR2017197} and IU-FSU \cite{Carbone} parameter sets. The nucleon mass ($M$) is 939.0 MeV.  All of the coupling parameters are dimensionless and the NM parameters are in MeV, except $k_3$ and $n_{0}$ which are in fm$^{-1}$ and fm$^{-3}$ respectively. The NM parameters are given at saturation point and T = 0 K for NL3, G3 and IU-FSU parameter sets in the lower panel. The references are $[a]$,$[b]$, $[c]$ $\&$ $[d]$ \cite{Zyla_2020}, $[e] $\&$ [f]$ \cite{Bethe_1971}, $[g]$ \cite{Garg_2018}, $[h] $\&$ [i]$ \cite{Danielewicz_2014}, and $[j]$ \cite{zimmerman}.}
\begin{tabular}{cccccccccc}
\hline
\hline
\multicolumn{1}{c}{Parameter}
&\multicolumn{1}{c}{NL3}
&\multicolumn{1}{c}{G3}
&\multicolumn{1}{c}{IU-FSU}
&\multicolumn{1}{c}{Empirical/Expt. Value}\\
\hline
$m_{\sigma}/M$ & 0.541 & 0.559 & 0.523 &0.426 -- 0.745 $[a]$\\
$m_{\omega}/M$  &  0.833  &  0.832 & 0.833 & 0.833 -- 0.834 $[b]$  \\
$m_{\rho}/M$  &  0.812  &  0.820 & 0.812 & 0.825 -- 0.826 $[c]$\\
$m_{\delta}/M$   & 0.0  &   1.043 & 0.0 & 1.022 -- 1.064 $[d]$\\
$g_{\sigma}/4 \pi$  &  0.813  &  0.782 & 0.793 & \\
$g_{\omega}/4 \pi$  &  1.024  &  0.923 & 1.037 &\\
$g_{\rho}/4 \pi$  &  0.712 &  0.962  & 1.081 & \\
$g_{\delta}/4 \pi$  &  0.0  &  0.160 & 0.0 & \\
$k_{3} $   &  1.465 &    2.606 & 1.1593 &  \\
$k_{4}$  &  -5.688  & 1.694   & 0.0966 &\\
$\zeta_{0}$  &  0.0 &  1.010 & 0.03 &   \\
$\eta_{1}$  &  0.0 &  0.424 & 0.0 & \\
$\eta_{2}$  &  0.0 &  0.114 & 0.0 &  \\
$\eta_{\rho}$  &  0.0 &  0.645 & 0.0 & \\
$\Lambda_{\omega}$  &  0.0 &  0.038 & 0.046 &  \\
\hline
$n_{0}$ & 0.148 & 0.148 & 0.154 & 0.148 -- 0.185 $[e]$\\
$B.E.$ & -16.29 & -16.02 & -16.39 & -15.00 -- 17.00 $[f]$\\
$K_{0}$ & 271.38 & 243.96 & 231.31 &220 -- 260 $[g]$\\
$F_{sym,0}$ & 37.43 & 31.84 & 32.71 & 30.20 -- 33.70 $[h]$\\
$L_{sym,0}$ & 120.65 & 49.31 & 49.26 & 35.00 -- 70.00 $[i]$ \\
$K_{sym,0}$ & 101.34 & -106.07 & 23.28 & -174 -- -31 $[j]$\\
$Q_{sym,0}$ & 177.90 & 915.47 & 536.46 & -----------\\
\hline
\hline
\end{tabular}
\label{table1}
\end{table}
%
We start our discussion with the energy and pressure of hot NM and procurement of critical temperature for liquid-gas phase transition. In Fig.\ref{EOS}, we depicted the calculated results of binding energy per nucleon and pressure density from $T = 0$ to $20$ MeV as a function of density. We observed an increase in the binding energy and saturation density of the SNM with temperature for all the three supposed parameter sets. The marked point on each curve of the upper panel of the Fig.\ref{EOS} represents its minima. It has been already reported that NM saturates around $0.148$ fm$^{-3}$ with binding energy per nucleon around $-16$ MeV for both the parameter sets (NL3 and G3) at zero temperature \cite{Bharat} while the saturation density for IU-FSU parameter set is 0.154 fm$^{-3}$. Here we found that as we increase the temperature from $0$ to $20$ MeV, $n_0$ increases linearly from $0.148$ to $0.196$ fm$^{-3}$ and similarly the binding energy also goes on increasing which is clear from the Fig.\ref{EOS}. The increase in the binding energy per nucleon with temperature indicates that the system becomes more loosely bound at the higher temperature.
%
The lower panel of Fig.\ref{EOS} shows the variation of pressure with temperature as a function of density. This adaptation of pressure is utterly important in determining the critical parameters of the liquid-gas phase transition, especially the critical temperature, $T_{C}$. Various theoretical and experimental studies predict the value of $T_{C}$ for SNM in the range of $10-20~$MeV \cite{SubrataPal,KUPPER,BoNan,Moretto,Tatsuyuki}. We also found the value of $T_C$ using equation \ref{critical} as $14.60$, $15.37$ and $14.50$ MeV for NL3, G3 and IU-FSU parameter sets respectively. Other important characteristics quantities of the liquid-gas phase transition like $P_C, \rho_C$ can also be determined using $T_C$ with the help of correlations derived in reference \cite{BoNan}. 
\begin{figure}[H]
    \centering
    \includegraphics[width=0.8\columnwidth]{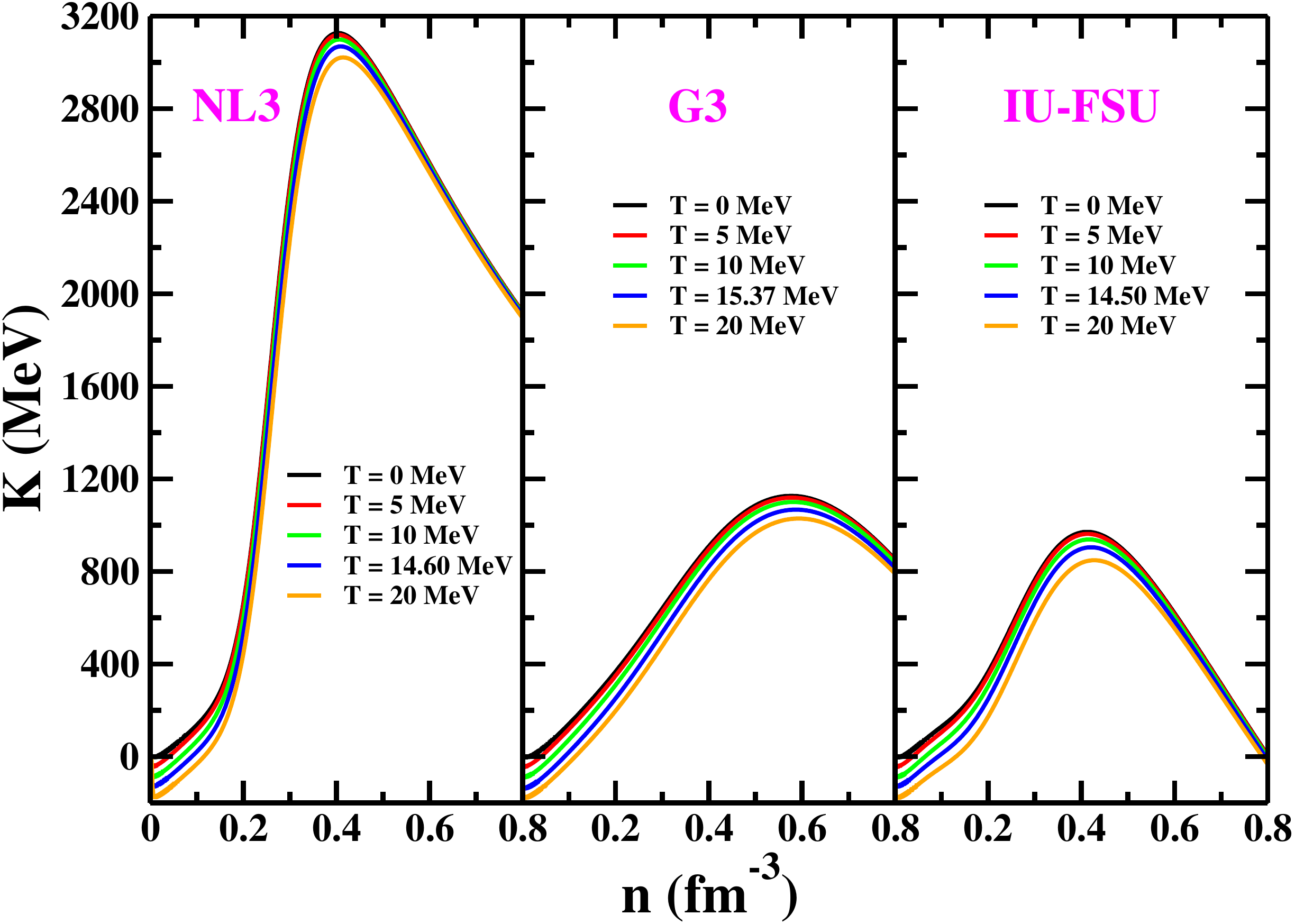}
    \caption{(colour online) Incompressibility $(K)$ as a function of nucleon density for SNM ($t=0$). The left panel shows the results for NL3, middle panel for G3 and right panel for IU-FSU parameter set.}
    \label{incomp}
\end{figure}
The variation of the incompressibility modulus has also been depicted in Fig. \ref{incomp} and \ref{incompvari}. As the value of n increases, the chart of $K$ exhibits anomalous behaviour. We observed that the magnitude of $K$ shows a maximum around $4-5$ times of the saturation density for all the three parameter sets. However, the magnitude of $K$ is quite high for NL3 parameter set in comparison to G3 and IU-FSU, which indicates a strong dependence of $K$ on the nature of the EoS. We also observed an unfamiliar proneness in the K values at saturation density (denoted as $K_0$) for different temperature which are shown in Fig. \ref{incompvari}.For NL3 parameter set $K_0$ shows an increment with increase in temperature while for G3 and IU-FSU parameter sets the case is reversed. This contemplation of $K_0$ confirms the sensitiveness of incompressibility on the choice of parametrization and the cross-coupling of field variables responsible for the softening of EoS. This fact has also been supported by the study in reference \cite{Mekjian}.
We also analysed that the G3 and IUFSU parameter sets indulge all the constraints of the elliptic flow heavy-ion collision experiment \cite{Chenchen} on the value of $K_0$, which is $K_0$ = $220\pm40$ MeV. NL3, on the other hand, being the stiff parameter set depicts the higher value of $K_0$ and does not support the experimental results. The effect of the temperature on free symmetry energy and its derivatives has also been studied from lower to higher density of NM. Fig. \ref{symm} reflects the effect of temperature on $F_{sym}$ and its first derivative $L_{sym}$ (i.e slope parameter) of SNM. The rise in temperature does not hint any significant change in the symmetry energy of the system at high density. However, the magnitude of the symmetry energy at saturation density ($F_{sym,0}$) increases as we increase the temperature. Another important aspect that we observed here is the difference in the magnitude of the $F_{sym}$ for NL3 and G3 parameter sets. The magnitude range of $F_{sym}$ for NL3 is quite high in comparison to G3 and IU-FSU, which indicates that the stiff EoS allocate the higher value of symmetry energy. 
%
Again the determination of the higher value of $F_{sym,0}$ along with $K_0$ by NL3 forces suggested that we need a better parameterized EoS, so that we can match the prediction of nuclear parameters made by various theoretical and experimental studies. On the other hand, the spectrum of $F_{sym}$ at finite temperature determined by our recently developed G3 parameter and the popular IU-FSU, fall in the expected range which is verified by other studies also \cite{Yingxun, Agrawal_2014, Baldo, Colonna, Carbone}. $L_{sym}$ changes slightly as we change the temperature for NL3 parameter set, while for G3 and IU-FSU it remains unaffected. The change in slope parameter at saturation density ($L_{sym,0}$) with temperature is shown in Fig. \ref{symmvari}. The effect of temperature on the second and third derivatives of $F_{sym}$, which are known as curvature and isovector skewness parameter respectively (denoted by $K_{sym}$ and $Q_{sym}$) is shown in Fig. \ref{Q_0} and \ref{Qvari} for both the parameter sets.

The variation of $K_{sym}$ and $Q_{sym}$ with density somehow reflects the sight of the sinusoidal wave, with extremum of the curve near twice the saturation density as can be seen in Fig. \ref{Q_0}. $K_{sym}$, being the higher order parameter, can be constrained with the help of neutron star observations \cite{zimmerman}. Recently, a study done by Josef Zimmerman et al. merged the data reported by two most vital experiments (PSR J0030+0451 by NICER Collaboration \cite{Riley_2019} and GW170817 by LIGO/Virgo \cite{PhysRevLett.119.161101, Abbott_2018}) and derive a joint 1-$\sigma$ constraint on curvature parameter at saturation density ($K_{sym,0}$). It is believed to be the most reliable bound on $K_{sym,0}$ till date and reported within the 90 \% confident bounds as $K_{sym,0} = -102^{+71}_{-72}$ MeV \cite{zimmerman}.
%
%
In Fig. \ref{Qvari} we can see that the value of $K_{sym,0}$ estimated by G3 parameter set for entire temperature spectrum falls into the range of 1-$\sigma$ constraint discussed above. NL3 and IU-FSU parameter sets, on the other hand, not only inconsistent with the estimated constraint but surprisingly falls entirely in the opposite magnitude range. This observation clearly rejects the NL3 and IU-FSU phenomenological model in determining the NM parameters for astrophysical observations and hints for G3 EoS, best suited for the study of astrophysical phenomena. Studies show that there is a strong correlation between $K_{sym,0}$, tidal deformability and $R_{1.4}$ (radius of NS with mass $1.4$ $M_{\odot}$) \cite{Fortin, Steiner, Yagi}. By obligating $K_{sym,0}$ in the befitting range, we can measure tidal deformability theoretically, which is quite a complicated quantity to measure independently, precisely up to a certain level of accuracy. $Q_{sym}$ being the most ambiguous quantity also shows entirely different behaviour for the NL3 and G3 parameter set. For G3 it falls in the positive magnitude range while for NL3 and IU-FSU, it beholds the negative magnitude as the temperature increases. Although there is no reported constraint on $Q_{sym,0}$ till now, but we believe that since G3 parameter set satisfy all the desired results for other NM parameters, so it is obvious to assume it of the right kind. However, some predictions made on the basis of skyrme interactions supports the negative magnitude of skewness parameter \cite{Tews_2017}, so it is difficult to state anything about $Q_{sym}$ with certainty. 
\begin{figure}[H]
    \centering
    \includegraphics[width=0.8\columnwidth]{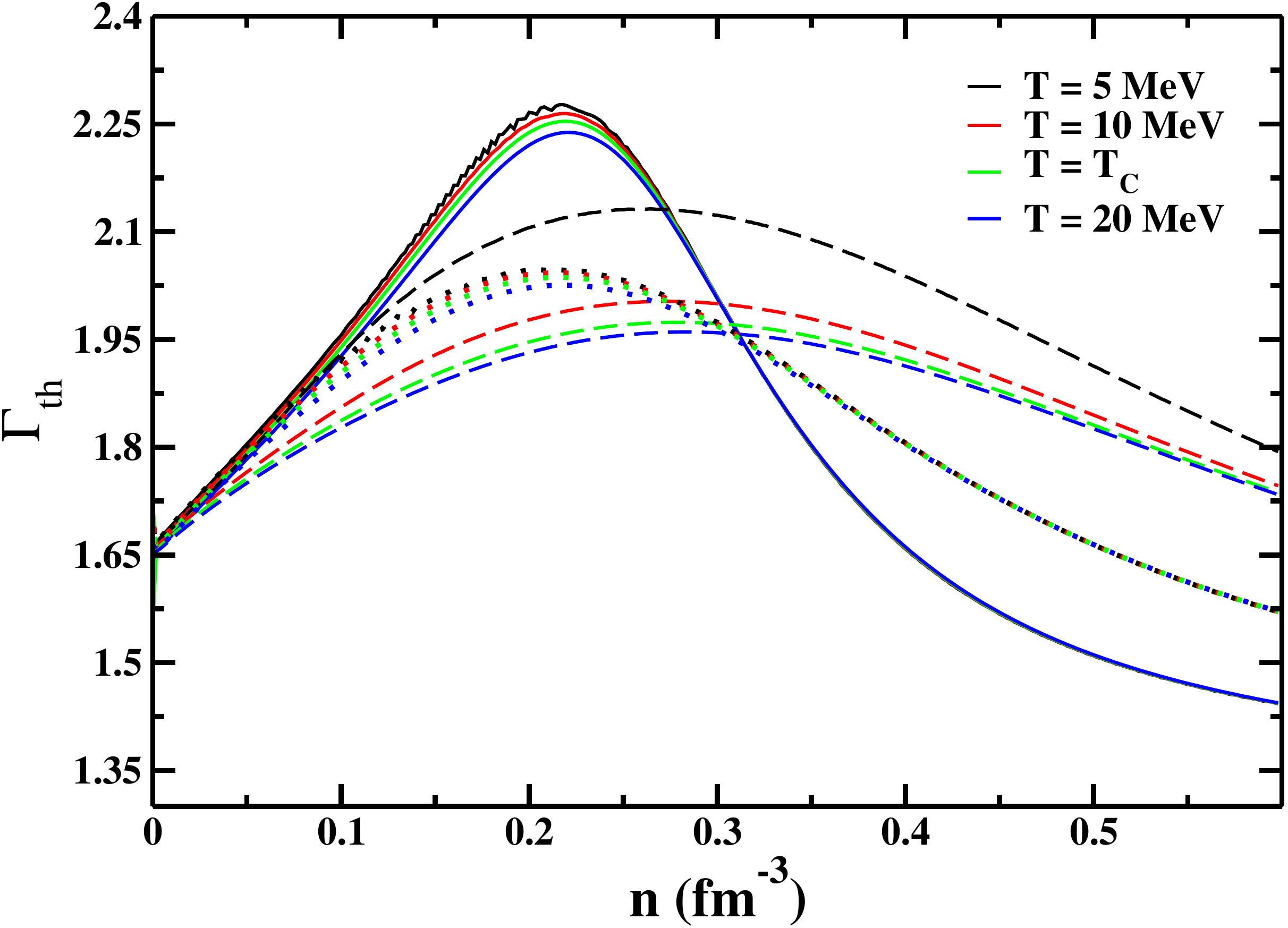}
    \caption{(colour online) The thermal index at different temperatures as a function of nucleon density for NL3 (solid line), G3 (dashed line) and IU-FSU (dotted) parameter sets. $T_C$ is the critical temperature for the corresponding parameter set}
    \label{thermal}
\end{figure}
Fig. \ref{thermal} enounce the variation of the thermal index for the avowed parameter sets. The nature of the EoS of the astrophysical phenomenon is also dictated in some manner by the density dependence of the thermal index. As we can see in Fig. \ref{thermal} that the thermal index in the low density region approaches the value of 5/3 for all temperatures, which indicates that the system described by our model for all the parameter sets almost behaves like a non-relativistic ideal gas \cite{Carbone}. We also observe that as the temperature goes on increasing the maximum value of the thermal index decrease, which is consistent with the fact that at high temperature, the system approaches relativistic behaviour. The value of $\Gamma_{th}$ around 2 has been appraised as an important aspect in the neutron-star merger simulations, which has also been reflected in our calculations \cite{Bauswein}. The thermal index curve for NL3 is comparatively steeper than G3 and IU-FSU parameter sets and approaches towards the relativistic gas behaviour more rapidly. The variations in the thermal index with density for different EoS solidify the frequency of gravitational wave oscillations and also decides the delay time of black hole collapse \cite{Bauswein,Carbone}. 
\section{Cooling through neutrino emission}\label{cooling}
In this section, we provide an approach to study the thermal evolution of the newly born neutron star through EOS. It is widely believed that the dense neutron star formed just after the supernova explosion contains tantamount nucleons, i.e. almost the equal number of protons and neutrons \cite{Lattimer}. After that, the process of cooling and neutronization \cite{Lattimer,stgaard} happens slowly, and finally, it achieves the thermal stabilization and the beta equilibrium, which we call cold neutron star. The direct Urca processes $n \longrightarrow p + e^- + \bar\nu_e$ and $p + e^- \longrightarrow n + \nu_e$, governs the cooling of newly born neutron stars \cite{Boguta} through neutrino emission and requires a super threshold proton fraction inside the super-dense star core to operate. To realize the effect of neutrino emissivity on cooling mechanism and EOS, we consider a degenerate dense matter containing nucleons (neutrons and protons) and the electrons at finite temperature, for which the Lagrangian is redefined as
%
\begin{eqnarray}
{\cal L}_{total} &=& {\cal L} + \bar\phi\,(i\gamma_{\mu} \partial^{\mu} - m_e)\phi,
\label{Ltot}
\end{eqnarray}
%
where ${\cal L}$ is the Lagrangian defined in Eq. (\ref{lag}), $m_e$ being the mass of the electron and the last term represents the electrons contribution in the matter, which are considered to be non-interacting particles. Following the same procedure as defined in section \ref{QHD}, we will get the total energy density and pressure for this system
%
\begin{eqnarray}\label{edt}
 E_{total} &=&  E + \frac{2}{(2\pi)^{3}}\int_{0}^{k_{e}} d^{3}k\, {\cal E}_{e} (k)\, [f_{e}(T)+\bar f_{e}(T) ], \\
 P_{total} &=&  P + \frac{2}{3(2\pi)^{3}}\int_{0}^{k_{e}} d^{3}k\, \frac{k^2}{{\cal E}_{e}}\, \Big[f_{e}(T)+\bar f_{e}(T)\Big] , 
\end{eqnarray}
%
where E and P is the energy density and pressure defined in equation \ref{ed}, $k_e$ is the Fermi momentum of the electron, $f_e$ is the Fermi distribution of electron for finite temperature and ${\cal E}_e$ is the energy of electron given by
\begin{eqnarray}
{\cal E}_e &=& \sqrt{k^2_e + m^2_e}.
\end{eqnarray} 
We maintained the same number density for proton and electron, i.e. $n_p=n_e$, to achieve charge neutrality in the described system.\\

The expression for the neutrino emissivity ($Q$) in high dense (neutron star) system was first estimated by Lattimer et al. in non-relativistic manner \cite{LattimerUrca}. Since the movement of nucleons in neutron star cores is relativistic, so the relativistic expression for neutrino emissivity was later calculated by L. B. Leinson and A. Perez. The non-relativistic emissivity is quite small than what is predicted by the relativistic formalism. The detailed explanations and the derivation for the formula of neutrino emissivity ($Q$) in the relativistic framework and mean-field approximation, which is used here can be found in the reference \cite{Leinson_2001,Leinson_2002}.

\begin{figure}[H]
    \centering
    \includegraphics[width=0.8\columnwidth]{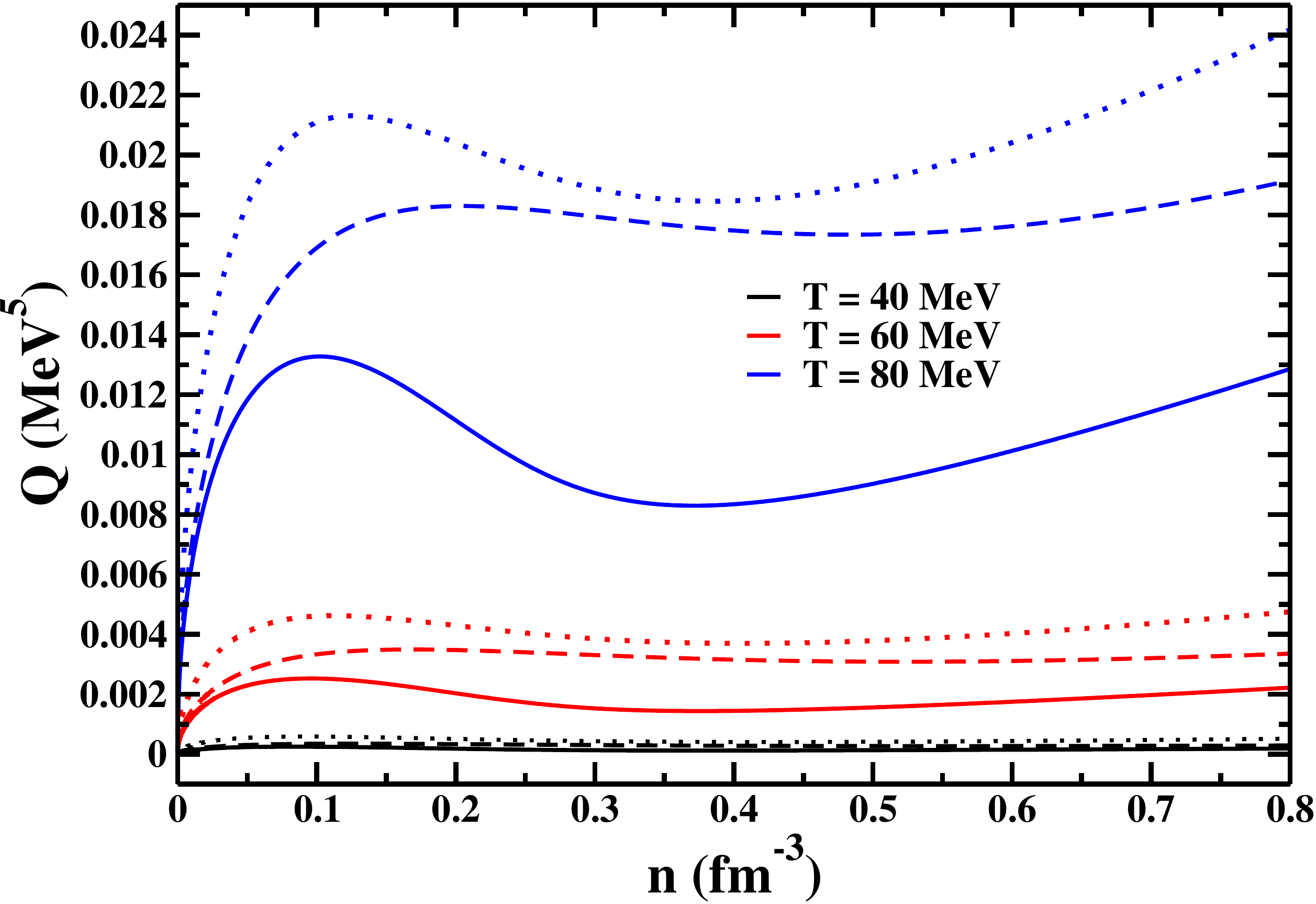}
    \caption{(colour online) Neutrino emissivity at different temperatures as a function of nucleon density for NL3 (solid line), G3 (dashed line) and IU-FSU (dotted line) parameter sets.}
    \label{coolingraph}
\end{figure}

The formula for neutrino emissivity is given by \cite{Leinson_2002}
\begin{eqnarray}
Q &=& \frac{457\pi}{10080} G_F^{2}\,C^2\,T^6\,\Theta(k_e + k_p - k_n) \Bigg\{\left(C_A^2-C_V^2\right)  \nonumber\\
&&
M_p^\ast\, M_n^\ast\, {\cal E}_e+ \frac{1}{2} \left(C_V^2+C_A^2\right) \bigg[4\,{\cal E}_n\, {\cal E}_p\,  {\cal E}_e - \left({\cal E}_n - {\cal E}_p\right) \nonumber  \\ 
&& 
\left(\left({\cal E}_n + {\cal E}_p \right)^2 - k_e^2\right) \bigg] + C_V\, C_M\, \frac{\sqrt{M_p^\ast\,M_n^\ast}} {M}\nonumber  \\ 
&& 
\left[2 \left({\cal E}_n - {\cal E}_p \right) k_e^2 - \left(3 \left(  {\cal E}_n - {\cal E}_p \right)^2 - k_e^2 \right){\cal E}_e \right] 
\nonumber\\
&&  
+ C_A\left(C_V + 2\frac{\sqrt{M_p^\ast M_n^\ast}}{M} C_M \right) \left({\cal E}_n + {\cal E}_p \right) \nonumber \\ 
&& 
\left(  k_e^2 - \left({\cal E}_n + {\cal E}_p \right)^2\right)
+ C_M^2 \frac{1}{4M^2} \bigg[8M^{\ast 2} \left( {\cal E}_n - {\cal E}_p \right) \nonumber  \\ 
&& 
\left(k_e^2 - \left( {\cal E}_n - {\cal E}_p\right) {\cal E}_e \right)
+ \left(k_e^2 - \left( {\cal E}_n - {\cal E}_p\right)^2 \right)\nonumber\\ 
&& 
\left(2{\cal E}_n^2 
+ 2{\cal E}_p^2 - k_e^2 \right) {\cal E}_e - \left(k_e^2 - \left( {\cal E}_n - {\cal E}_p\right)^2 \right) \nonumber\\ 
&& 
\left({\cal E}_n + {\cal E}_p \right)\left(2{\cal E}_n - 2{\cal E}_p - {\cal E}_e \right)  \bigg] \Bigg\}, 
\end{eqnarray}
where $G_F = 1.166\times10^{-11}$ MeV$^{-2}$ is the Fermi weak coupling constant, $C$ = 0.973 is the Cabibbo factor, $C_V$ = 1 and $C_A$ = 1.26 are the vector and axial-vector constants respectively and the constant $C_M$ = 3.7 represents the weak magnetism effects. The condition inevitable for Urca processes to go is represented by $\Theta(k_e + k_p - k_n)$ = 1, if $k_e + k_p - k_n \geq 0$ and zero otherwise, where $k_e$, $k_p$ and $k_n$ are the Fermi momenta of electron, proton and neutron respectively.\\
Since our defined system of nucleons and leptons (n-p-e) clearly satisfy the above necessary condition required for the initialisation of direct Urca process, so, we calculated the neutrino emissivity for both NL3, G3 and IU-FSU parameter sets which is displayed in Fig. \ref{coolingraph}. The dashed lines in Fig. \ref{coolingraph} represent the results for G3, the solid lines stand for NL3 and the dotted lines for IU-FSU parameter set. Some very interesting remarks can be concluded based on these curves of $Q$ depicted at different temperatures. The most vital observation of this work is that the neutrino emissivity is responsible for the cooling of newly born neutron star only in its initial stage, i.e. when the temperature is quite high, and the neutrons are enough thermally excited to trigger the direct Urca process. As the star cools down, the magnitude of Q decreases substantially and then cooling mainly takes place through photon emission. This behaviour can be seen clearly in Fig. \ref{coolingraph} for all the assumed parameter sets. The magnitude of $Q$ for 40 MeV (represented by the black line) temperature is negligible in comparison to 80 MeV curve (blue line).   
Also, initially, the diffusion rate of neutrinos is so high that the matter cools down within a fraction of seconds and it is believed that it is so effective that it lowers the temperature to about 1 billion Kelvin during this momentary span \cite{Dmitry,Cumming}. Another important dimension that we observe in this curve is the difference in the magnitude of the $Q$ values for NL3, G3 and IU-FSU parameter sets. This tells us about the dependence of cooling property of newly born NS on the EoS. As we know that NL3 provides the stiffest EoS, which means the high mass of the proto-neutron star and G3 being the softer EoS in comparison to NL3 predicts lower mass. Keeping this fact in mind, we concluded that the cooling of proto-neutron star through direct Urca process is slow in heavier star and fast in the lighter one. 
\section{Proto-Neutron Star}\label{star}
\begin{figure}[H]
    \centering
    \includegraphics[width=0.8\columnwidth]{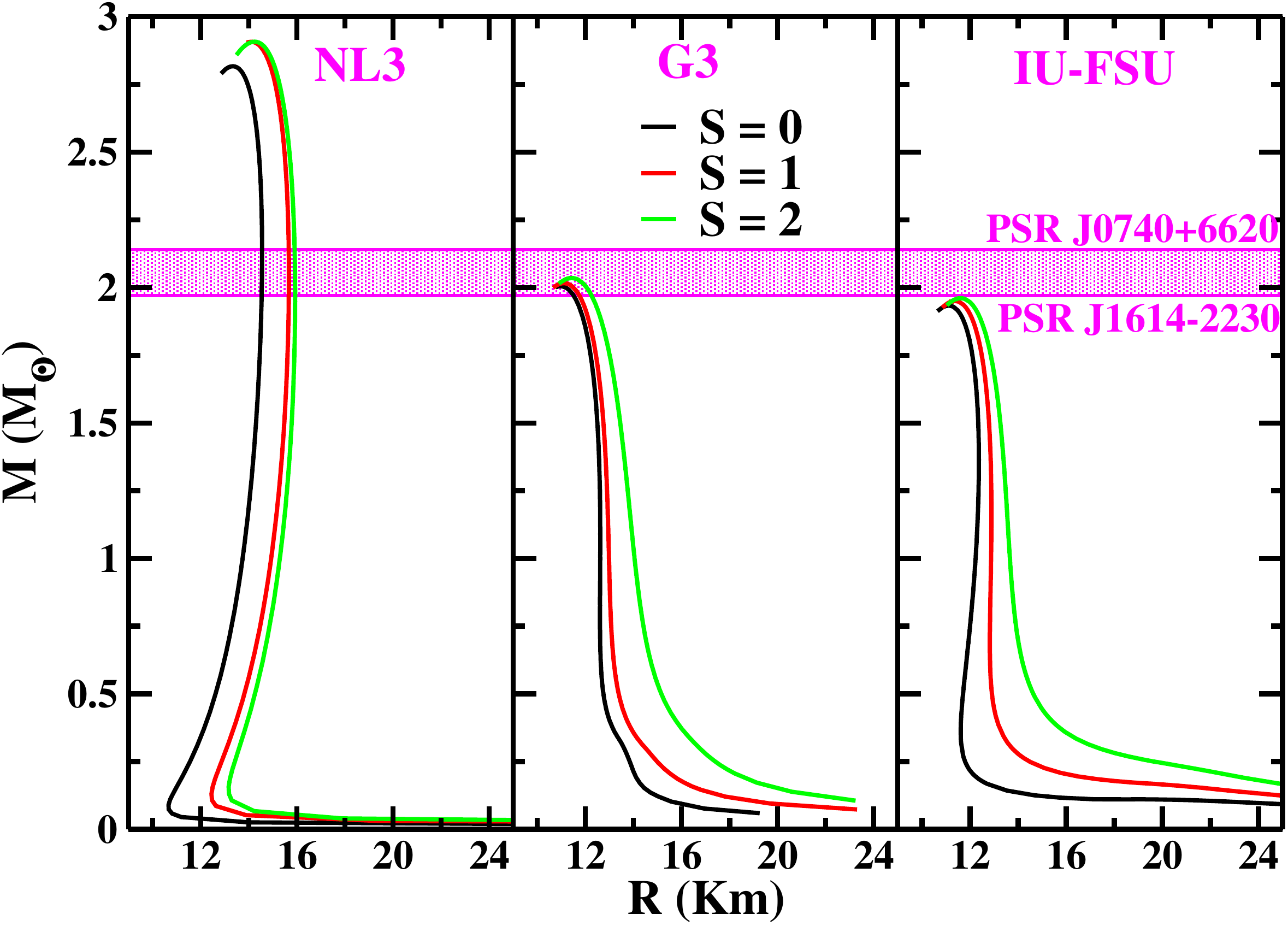}
    \caption{(colour online) M-R profile of proto-neutron star at different entropy for NL3 (left panel), G3 (middle) and IU-FSU (right panel) parameter sets respectively.}
    \label{M-R-C}
\end{figure}

To understand the complete penchant of temperature on a dense matter system, we extend our calculations to obtain the mass-radius (M-R) profile of proto-neutron star (PNS) in this section. The newly born proto-neutron star is hot, lepton-rich and has a core in the evolutionary stage which ultimately grow either in a cold neutron star or crash into a black hole. As we discussed in the previous section that a PNS losses a significant amount of energy through neutrino emission and neutrino emissivity plays an important role in the evolution of the newly born star. However, some of the neutrinos with small mean free path unable to escape the core of the neutron star and got trapped inside the star. These trapped neutrinos affect the early evolutionary illation and properties of the PNS \cite{camelio2018, HuanyuandHong, Pons_1999}. We modify our Lagrangian mentioned in Eq. (\ref{Ltot}) and added the terms responsible for neutrino effects to derive the EoS for the proto-neutron star system envisaged in the present section. The modified Lagrangian is
\begin{eqnarray}
{\cal L}_{PNS} &=& {\cal L}_{total} + \bar\phi_{\nu_e}\,(i\gamma_{\mu} \partial^{\mu})\phi_{\nu_e},
\label{LPNS}
\end{eqnarray}
where $\phi_{\nu_e}$ is the wave-function of electron neutrino. The relevant energy density and pressure for the above mentioned Lagrangian can be derived as \cite{HuanyuandHong}
\begin{eqnarray} \label{EoSS}
 E_{PNS} &=&  E_{total} + \sum_{\nu_e} \Bigg(\frac{7\pi^2}{120} + \frac{\mu^2_{\nu_e}T^2}{12} - \frac{\mu^4_{\nu_e}}{24 \pi^2} \Bigg), \\
 P_{PNS} &=& P_{total} + \sum_{\nu_e} \frac{1}{360} \Bigg( 7\pi^2 T^{4} + 30 \mu^2_{\nu_e} T^2 + \frac{15 \mu^4_{\nu_e}}{\pi^2} \Bigg)
\end{eqnarray}
where the summation is over the total number of trapped neutrinos; $E_{total}$ and $P_{total}$ are the energy density and pressure defined in eq. \ref{edt} and $\mu_{\nu_e}$ is the chemical potential of electron neutrino. We also maintained the necessary $\beta-$ equilibrium and charge neutrality conditions, i.e.
\begin{eqnarray}
\mu_{n} &=& \mu_{p} + (\mu_{e} - \mu_{\nu_e}), \\ \nonumber
n_{p} &=& n_{e},
\end{eqnarray}
where $\mu_{n}, \mu_{p}, \mu_{e}$  and $\mu_{\nu_e}$ are the chemical potentials of the neutron, proton, electron and neutrino respectively; $n_{p}$ and $n_{e}$ are the number densities of proton and electron. We can also define the lepton fraction as  
\begin{eqnarray}
Y_{L} &=& \frac{n_{e} + n_{\nu_e}}{n}
\end{eqnarray}
where $n_{e}$, $n_{\nu_e}$ and n are the number density of electrons, neutrinos and baryons respectively. The effect of trapped neutrinos on the early evolutionary stages of PNSs can be represented by fixing $Y_{L}$ in a certain defined range, i.e. $Y_{L} \sim 0.1-0.4$. We fix $Y_{L} = 0.4$ in our calculations, which is the best precedence to explore the properties of newly born PNS \cite{HuanyuandHong, Prakash_1997, Pons_1999}. Also, the entropy per baryon ($S$) of the star matter can be derived as \cite{Prakash_1997} 
\begin{eqnarray}
S &=& \frac{E_{PNS} + P_{PNS} - \sum_{\alpha=p,n} n_{\alpha} \mu_{\alpha}}{nT}
\end{eqnarray}
For a fixed entropy, the effect of the neutrino trapping is to keep the electrons concentration high so that matter is more proton-rich in comparison to the case in which the neutrinos are not trapped \cite{Prakash}. We study the M-R profile for the PNS using a different kind of EoS, first by placing the temperature constant throughout the star (fixed temperature EoS) \cite{MESQUITA, Dexheimer_2008, Manka_2001} and another by fixing the certain defined values of entropy per baryon ($S=1\&2$) for the star matter (fixed entropy EoS) \cite{Burgio, BIN2016340, Hong_2016, Strobel_2001}. Although we have used the constant temperature EoS to describe the mass-radius profile of proto-neutron star, but this assumption is valid only for low temperature spectrum. If the matter's temperature is significantly larger than the corresponding critical Fermi temperature, the star becomes unstable. Also, the temperature of the star varies significantly from core to the surface of the star. So, to debar all these restrictions and for a better undisputed M-R outline of PNS, we extended our calculations for fixed entropy EoS. The method of fixed entropy EoS is more appropriate to study the properties of PNS at finite temperature.\\
By imposing the above stated conditions on the described Lagrangian, we can easily calculate the mass and radius of the static isotropic proto-neutron star for both kind of EoSs using the Tolman-Oppenheimer-Volkov (TOV) equations \cite{tolman,Oppenheimer}. The TOV equations are
\begin{eqnarray}
\frac{d P_{PNS}(r)}{d r} &=& -\frac{[E_{PNS}(r)+P_{PNS}(r)][M(r)+{4\pi r^3 P_{PNS}(r)}]}{r^2\Big(1-\frac{2M(r)}{ r}\Big)}, \\ \nonumber \\
\frac{d M(r)}{d r} &=& 4\pi r^2 {E_{PNS}(r)}.
\end{eqnarray}
Where $E_{PNS}$ and $P_{PNS}$ are the energy density and pressure defined in eqs. (38) and (39). These equations are integrated from $r = 0$ to the stellar surface $r = R$, where $P_{PNS}(R) = 0$, for a particular choice of central density $\rho_c = \rho(0)$ to determine the NS mass $M = M(R)$. The value of $\rho_c$ that produces the maximum mass $M_{max}$ for a given EoS is $\rho_{max}$. The crust part of the neutron star also plays an important role in determining the properties of the proto-neutron star \cite{Strobel_1999}. We have also used the crust EoS of the corresponding temperature for a complete contact of Mass-Radius profile \cite{Togashi_2017}.
\begin{figure}[H]
    \centering
    \includegraphics[width=0.8\columnwidth]{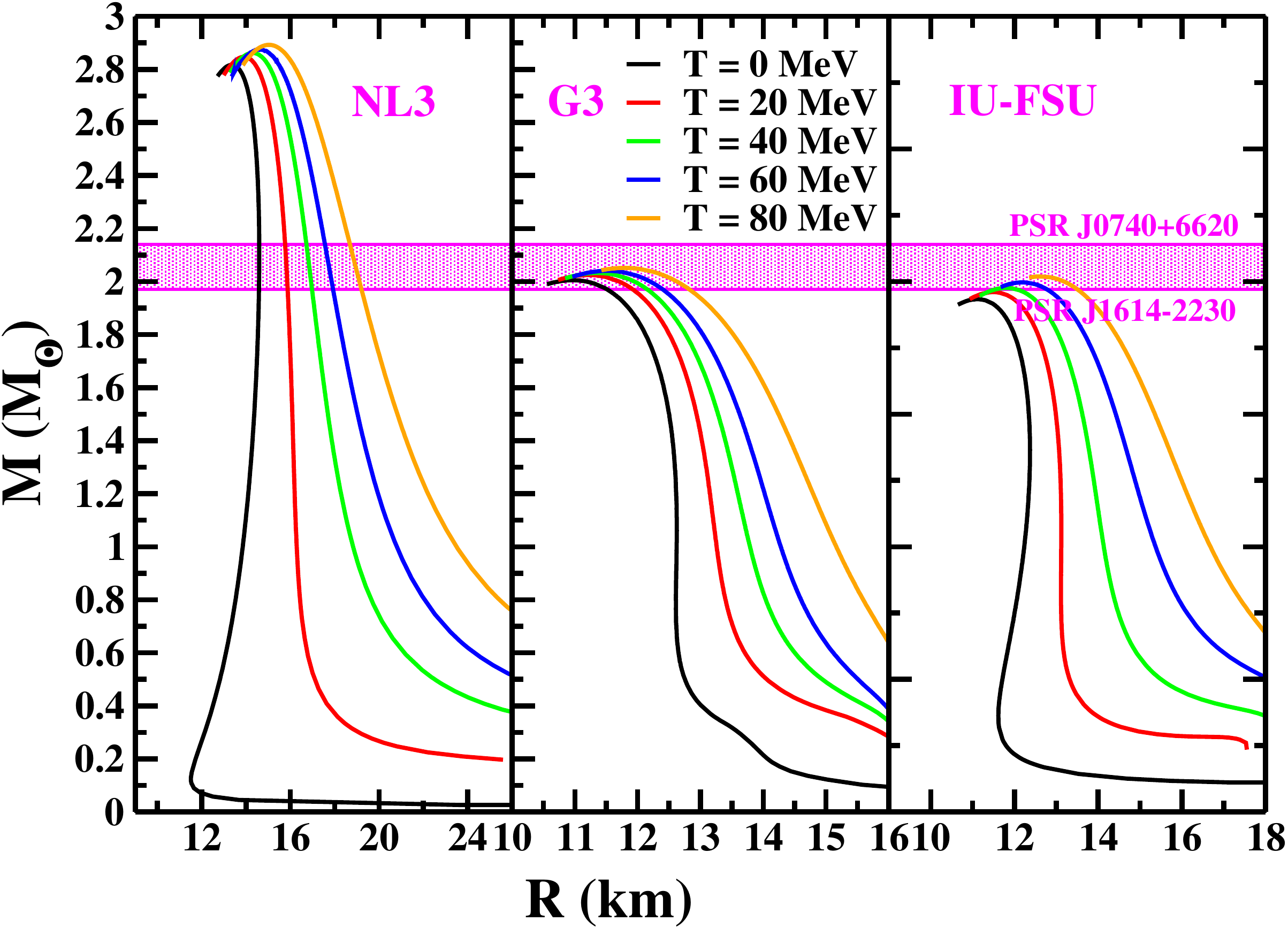}
    \caption{(colour online) M-R profile of proto-neutron star at different temperatures for NL3 (left panel), G3 (middle) and IUFSU (right panel) parameter sets respectively.}
    \label{M-R}
\end{figure}
The mass-radius profile of the  proto-neutron star for fixed temperature EoS along with trapped neutrinos is plotted in Fig.\ref{M-R}. The results from the precisely measured neutron stars masses, such as PSR J1614-2230 with mass $M=1.97\pm0.04M_\odot$ \cite{Demorest_2010} and PSR J0740+6620 with $M=2.15^{+0.10}_{-0.09}M_\odot$ \cite{Cromartie_2019} are shown in the horizontal bars in pink color. These observations suggest that the maximum mass predicted by any theoretical model should reach the limit of $\sim2.0M_\odot$, and this condition is satisfied by all of the EoSs taken into consideration. However, G3 parameter set follows the observational constraint of PSR J0740+6620 for the entire assumptive temperature range. We observe that the inclusion of neutrino trapping flattens the M-R curve at the top for G3 and IUFSU parameter sets, which affect the radius of the proto-neutron star considerably. We also notice that the inclusion of temperature increases the pressure at a given baryon density which yields the increase in the mass-radius profile of the proto-neutron star. The proto-neutron star has a little-bit large mass compared with that of the neutron star at zero temperature because the EoS is stiffer in the former case. The mass-radius profile of proto-neutron star using constant entropy EoS have also been studied and depicted in Fig. \ref{M-R-C}.
By analysing Fig. \ref{M-R} and \ref{M-R-C}, we concluded that the maximum mass determined by the G3 parameter set strongly satisfy the constraints of the PSR J0740+6620 and PSR J1614-2230. We observe an increase in maximum mass and radius in comparison to the cold NS for both kind of EoS (fixed T and S). The maximum mass, radius and central temperature ($T_{Ce}$) of the PNS for all the three parameter sets at different entropy is given in Table \ref{table2}. The maximum mass of the PNS for all the three parameter sets goes on increasing with increase in entropy of the system. Also, we observe that the massive PNSs have larger radius, so we concluded that more the entropy per baryon of the star, larger the radius of the system. In other words, we can say that the evolutionary process of the newly born star favours contraction in terms of size. It has also been reported that the maximum mass of a proto-neutron star is considerably affected by the earliest stage of its evolution, i.e. ratheripe type proto-neutron star ($\sim$ 1 sec after core bounce) have larger mass than the late-type proto-neutron star\cite{Strobel_2001}. So, the properties of the PNS depends strongly on how immaculately we gravitate the entropy per baryon of the system in its early stage.
\begin{table}
\centering
\caption{Maximum mass and radius of the PNS for NL3, G3 and IU-FSU parameter sets calculated using constant entropy EoS for S = 0, 1 and 2. $T_{Ce}$ denotes the central temperature of the PNS. }
\scalebox{0.7}{
\begin{tabular}{|l|lll|lll|lll|}
\hline
\multirow{2}{*}{\begin{tabular}[c]{@{}l@{}}Entropy\\  ($S$)\end{tabular}} & \multicolumn{3}{l|}{NL3} & \multicolumn{3}{l|}{G3} & \multicolumn{3}{l|}{IU-FSU} \\ \cline{2-10} & 
{\begin{tabular}[c]{@{}l@{}}$M$\\($M_\odot$)\end{tabular}}&
{\begin{tabular}[c]{@{}l@{}}$R$\\(km)\end{tabular}}&
{\begin{tabular}[c]{@{}l@{}}$T_{Ce}$\end{tabular}}&
{\begin{tabular}[c]{@{}l@{}}$M$\\($M_\odot$)\end{tabular}}&
{\begin{tabular}[c]{@{}l@{}}$R$\\(km)\end{tabular}}&
{\begin{tabular}[c]{@{}l@{}}$T_{Ce}$\end{tabular}}&
{\begin{tabular}[c]{@{}l@{}}$M$\\($M_\odot$)\end{tabular}}&
{\begin{tabular}[c]{@{}l@{}}$R$\\(km)\end{tabular}}&
{\begin{tabular}[c]{@{}l@{}}$T_{Ce}$\end{tabular}}
\\ \hline
 0&2.850& 13.32&0 & 2.004&10.95&0&1.930&11.12&0\\ \hline
 1& 2.905& 14.10&44.21 &2.016&11.07&40.38&1.950&11.37&39.91\\ \hline
 2& 2.908 & 14.34&82.24 & 2.034& 11.49&79.87& 1.960&11.63&78.85 \\ \hline
\end{tabular}}
\label{table2}
\end{table}
Also, to reconcile the results for maximum mass and radius determined by the two different approaches (i.e. constant entropy and constant temperature), we calculated the mass and radius of the PNS using constant temperature approach by fixing the temperature at $T = T_{Ce}$ for the corresponding parameter set. The results for the same have been presented in Table \ref{table3}.  We observe that the fixed temperature approach provide a larger mass and radius at $T=T_{Ce}$ of the corresponding entropy value. G3 parameter set demonstrates $\sim 2.5 \%$ and IU-FSU set shows $\sim 4.5 \%$ difference in the radius of the PNS calculated using constant entropy and temperature approach for the corresponding $T=T_{Ce}$. \\
\begin{table}
\centering
\caption{ Maximum mass and radius of the PNS for NL3, G3 and IU-FSU parameter sets calculated using constant temperature EoS. The central temperature at S = 0, 1 and 2 for the corresponding parameter set have been used as the constant temperature. }
\scalebox{0.8}{
\begin{tabular}{|l|ll|ll|ll|}
\hline
\multirow{2}{*}{\begin{tabular}[c]{@{}l@{}}Temp.\\  ($T$)\end{tabular}} & \multicolumn{2}{l|}{NL3} & \multicolumn{2}{l|}{G3} & \multicolumn{2}{l|}{IU-FSU} \\ \cline{2-7} & 
{\begin{tabular}[c]{@{}l@{}}$M$\\($M_\odot$)\end{tabular}}&
{\begin{tabular}[c]{@{}l@{}}$R$\\(km)\end{tabular}}&
{\begin{tabular}[c]{@{}l@{}}$M$\\($M_\odot$)\end{tabular}}&
{\begin{tabular}[c]{@{}l@{}}$R$\\(km)\end{tabular}}&
{\begin{tabular}[c]{@{}l@{}}$M$\\($M_\odot$)\end{tabular}}&
{\begin{tabular}[c]{@{}l@{}}$R$\\(km)\end{tabular}}
\\ \hline
$T_{Ce} (S=0)$&2.850&13.32&2.004&10.95&1.930&11.12\\ \hline
$T_{Ce} (S=1)$&2.880&14.29&2.035&11.36&1.987&11.89\\ \hline
$T_{Ce} (S=2)$&2.909&14.60&2.053&11.79&2.021&12.15 \\ \hline
\end{tabular}
}
\label{table3}
\end{table}
At last, we concluded that the properties of a PNS can be deciphered more appropriately by using the constant entropy perspective and as discussed in the previous sections also, since G3 parameter set respects all the experimental and observational constraints thoroughly, so it is one of the most compatible RMF parameter set.
\section{Summary and Conclusions}\label{discussion}
We have studied the consequences of finite temperature on the nuclear properties of SNM and the crucial section required for the cooling of remnants of supernovae explosion. We used well-known NL3 and IU-FSU and the recently developed G3 parameter sets of the RMF model for a comparative study. We concluded that NL3 with the possession of stiffest EoS does not provide the empirical values for most of the nuclear properties. However, IU-FSU satisfies most of the constraints on the NM properties but also neglect some of them. The variations of the binding energy and pressure with baryon density for different temperatures are qualitatively similar for all the three acquired parameter sets. However, G3 predicts the higher value of critical temperature for liquid-gas phase transition, which is more proximate to the reported experimental values. We also observed a contrarious development in the value of $K_0$ for the defined parameter sets with the increase of temperature. Both $K$ and $K_{sym}$ being the second derivatives of the different forms of energy, this behaviour of $K_0$ may be influenced by the magnitude of $K_{sym}$ which is positive for NL3 and negative for the G3 parameter set. $K_{sym}$ value of the G3 parameter set lies in the range reported by the NICER and LIGO collaboration. This result shows that the G3 parameter set is more suitable to reproduce the appropriate form of the EoS and can be used to study the properties of neutron stars more accurately. We did not observe any significant variation in the $Q_{sym}$ parameter at saturation density for the G3 parameter set, while its value decreases for the NL3 and IU-FSU parameter sets.\\
We emphasized on the cooling mechanism of highly dense matter and studied the effects of all three parameter sets on the neutrino emissivity. We used the relativistic approach to derive the detailed expression of neutrino emissivity, which is more effective than the non-relativistic approach. We concluded some important remarks about the cooling mechanism of the newly born proto-neutron star on the basis of the outcomes of our detailed calculation. We observed that the magnitude of neutrino emission is directly proportional to the temperature of the neutron star. As the body cools down, the magnitude of the neutrino emission decreases. Also, the neutrino emissivity has maximum value around the saturation density, which indicates that the saturated matter cools more rapidly through direct Urca process. Another important aspect that we inspect is that the neutrino emissivity is higher for that parameter set which provides softer EoS. As we can see in the previous section that the stiff EoS measures higher mass NS, so we concluded that the lighter remnant of the supernovae explosion cools down more expeditiously through neutrino emissivity of the direct Urca process. Moreover, We look forward to see if any post-merger signal is explored observationally in near future by LIGO
/VIRGO and NICER collaboration, which will help us to comprehend finite temperature NM physics more appropriately and extensively. We also take notice that the mass numerated by the G3 parameter set for the whole temperature range fits in the envelope determined by the GW170817 experimental data. Finally, we adduced that G3 parameter set is more appropriate to study the properties of stellar objects through EoS. \\
{\bf Acknowledgement}\\
 This work is partially supported by National Natural Science Foundation of China Grant No. 11873040.
\bibliographystyle{spphys}
\bibliography{ft.bib}
\end{document}